\documentclass[journal=ancac3,manuscript=article]{achemso}

\usepackage[artemisia]{textgreek}
\usepackage{gensymb}
\usepackage[version=3]{mhchem}
\usepackage{filecontents}

\usepackage{xr}

\makeatletter
\newcommand*{\addFileDependency}[1]{
  \typeout{(#1)}
  \@addtofilelist{#1}
  \IfFileExists{#1}{}{\typeout{No file #1.}}
}
\makeatother

\newcommand*{\myexternaldocument}[1]{%
    \externaldocument{#1}%
    \addFileDependency{#1.tex}%
    \addFileDependency{#1.aux}%
}

\myexternaldocument{SI}

\author{Daniel K. Angell}
\affiliation{Materials Science and Engineering, Stanford University, Stanford, CA, 94305, U.S.
}
\email{dkangell@stanford.edu}
\alsoaffiliation{contributed equally to this work}

\author{Shuo Li}
\affiliation{Materials Science and Engineering, Stanford University, Stanford, CA, 94305, U.S.
}
\alsoaffiliation{Stanford Institute for Materials and Energy Sciences, SLAC National Accelerator Laboratory, Menlo Park, 94025, California, USA }
\alsoaffiliation{contributed equally to this work}

\author{Hendrik Utzat}
\affiliation{Materials Science and Engineering, Stanford University, Stanford, CA, 94305, U.S.
}

\author{Matti L. S. Thurston}
\affiliation{Materials Science and Engineering, Stanford University, Stanford, CA, 94305, U.S.
}

\author{Yin Liu}
\affiliation{Materials Science and Engineering, Stanford University, Stanford, CA, 94305, U.S.
}

\author{Jeremy Dahl}
\affiliation{Stanford Institute for Materials and Energy Sciences, SLAC National Accelerator Laboratory, Menlo Park, 94025, California, USA }

\author{Robert Carlson}
\affiliation{Stanford Institute for Materials and Energy Sciences, SLAC National Accelerator Laboratory, Menlo Park, 94025, California, USA }

\author{Zhi-Xun Shen}
\affiliation{Departments of Physics and Applied Physics, Stanford University, Stanford, 94305, California, USA }
\alsoaffiliation{Stanford Institute for Materials and Energy Sciences, SLAC National Accelerator Laboratory, Menlo Park, 94025, California, USA }

\author{Robert Sinclair}
\affiliation{Materials Science and Engineering, Stanford University, Stanford, CA, 94305, U.S.
}

\author{Nicholas Melosh}
\affiliation{Materials Science and Engineering, Stanford University, Stanford, CA, 94305, U.S.
} 
\alsoaffiliation{Stanford Institute for Materials and Energy Sciences, SLAC National Accelerator Laboratory, Menlo Park, 94025, California, USA }

\author{Jennifer A. Dionne}
\affiliation{Materials Science and Engineering, Stanford University, Stanford, CA, 94305, U.S.
}
\email{jdionne@stanford.edu}

\title[An \textsf{achemso} demo]
  {Nanodiamond grain boundaries and lattice expansion drive Silicon vacancy emission heterogeneity}
  
\begin{document}


\begin{abstract}

Silicon-vacancy (SiV$^-$) centers in diamond are promising candidates as sources of single-photons in quantum networks due to their minimal phonon coupling and narrow optical linewidths. Correlating SiV$^-$ emission with the defect’s atomic-scale structure is important for controlling and optimizing quantum emission, but remains an outstanding challenge. Here, we use cathodoluminescence imaging in a scanning transmission electron microscope (STEM) to elucidate the structural sources of non-ideality in the SiV$^-$ emission from nanodiamonds with sub-nanometer-scale resolution. We show that different crystalline domains of a nanodiamond exhibit distinct zero-phonon line (ZPL) energies and differences in brightness, while near-surface SiV$^-$ emitters remain bright. We correlate these changes with local lattice expansion using 4D STEM and diffraction, and show that associated blue shifts from the ZPL are due to defect density heterogeneity, while red shifts are due to lattice distortions.

\end{abstract}

\newpage
\vspace{10pt}

Optically-addressable  defects in diamond have made significant application-driven inroads in quantum technologies~\cite{pingault2017coherent,muller2014optical,sukachev2017silicon,sipahigil2014indistinguishable}, sensors of stress and temperature,~\cite{choi_ultrasensitive_2019, broadway2019microscopic,hamlin2019extreme} and ensemble magnetometry~\cite{abraham2021nanotesla,taylor2008high}. The negatively charged silicon vacancy (SiV$^-$) defect in diamond has proven particularly promising for  applications requiring light-emission owing  to its high brightness~\cite{neu_single_2011}, near transform-limited linewidths~\cite{rogers2014multiple,li2016nonblinking,jantzen2016nanodiamonds}, and reduced electron-phonon coupling owing to its symmetric geometry and larger mass than the carbon atom~\cite{dietrich_isotopically_2014}. These properties also imbue single photons emitted from the SiV$^-$s with comparatively high degrees of indistinguishability, promising to drive optical quantum technologies~\cite{bradac_quantum_2019}. However, the still incomplete understanding of the photo-physics and its interrelation with the diamond atomic structure hampers further progress. For one, the SiV$^-$ defect has been associated with various spectrally distinct emission centers ranging from 700-800nm~\cite{lindner_strongly_2018}. These emitters have been associated with crystal quality within the diamond, and possibly with hydrogen interstitials~\cite{bray_localization_2016}. Additionally, like most point defects in diamond, the SiV$^-$ is prone to inhomogeneous broadening due to a number of factors. Stress within the diamond lattice can alter the optical properties of the SiV$^-$ defect itself~\cite{sternschulte_1681-ev_1994,sternschulte1995uniaxial}. By breaking defect \textit{$D_{3d}$} symmetry, diamond-internal strain can change the zero-phonon line (ZPL) energy as well as the splitting of ground and excited states~\cite{meesala_strain_2018}. Bulk crystal quality, such as consistent $\it{sp3}$ bonding, has also proven important for the reduction of inhomogeneous broadening~\cite{levchenko2015inhomogeneous}. Shallow SiV$^-$ defects in particular are affected by poor crystal surface $\it{sp3}$-bonding saturation~\cite{lang_long_2020,bradac_observation_2010,hertkorn_vacancy_2019}. While typically poor photon outcoupling from the optically high-index diamond host matrices can be overcome in diamond nanostructures, dimensionality reduction often exacerbates the intrinsic problem of spectral inhomogeneity~\cite{castelletto2011radiative}. Together with the intrinsically low emission quantum efficiency, the trade-off between spectral stability and brightness presents a significant challenge.

In order to study how SiV$^-$ defects behave when close to boundaries, crystal defects, or other potential recombination centers, it is imperative to measure their optical properties at their native lengthscales. Most previous studies of SiV$^-$s have used optical microscopy with insufficient spatial resolution to delineate the relationships between optical emission and the diamond crystal structure. Scanning probe microscopy (SPM) techniques, such as scanning tunneling microscopy (STM) and Kelvin probe force microscopy (KPFM), have proven useful for identifying and characterizing sub-surface NV$^-$ defects in the nearfield~\cite{pawlak2013local,yung2013tip}; however, these techniques rely on defect localization within the near-field of a scanning tip. Optical super-resolution techniques have been used to image defects in bulk diamonds. As such, charge-state depletion (CSD) microscopy (4.1 nm resolution)~\cite{chen2015subdiffraction}, and stimulated emission depletion (STED) microscopy (5 nm - 2 Å)~\cite{arroyo2013stimulated,silani2019stimulated,blom2017stimulated,rittweger2009sted,chen2013wide,silani2019stimulated} can identify defect emission with impressive spatial resolution. Additionally,  scanning transmission electron microscopy (STEM) combined with cathodoluminescence (CL) spectroscopy has also granted nanometer scale resolution in a sample’s optical properties, revealing heterogeneity in optical emission from a subwavelength volume~\cite{tizei_spectrally_2012,kociak_cathodoluminescence_2017,zhang2014silicon,tizei_spatially_2013}. To date, however, no experiments have provided spectroscopic analysis concurrent with local crystal structure (\textit{e.g.} identification of crystal stacking faults, twin planes, or non-diamond phases) to explain such heterogeneity. This lack of combined optical/structural readout has precluded the assessment of structural sources of single-emitter heterogeneity.

Among techniques that grant nanoscale spatial resolution, STEM-CL can be directly correlated with a multitude of structural and spectroscopic STEM/TEM techniques (\textit{e.g.} Electron Energy Loss Spectroscopy (EELS), nanobeam diffraction, holography, and high-resolution imaging). These capabilities make STEM-CL an ideal technique to study how SiV$^-$ optical properties change throughout the diamond host lattice, with potentially sub-angstrom resolution. Here, we use STEM-CL to delineate the multiple factors contributing to optical heterogeneity of SiV$^-$s in high-quality epitaxially-grown chemical vapor deposited (CVD) nanodiamonds. Spectral analysis of spatial CL maps demonstrates that unstable emission centers associated with Si incorporation are spatially located at 2D defects such as at grain boundaries within the nanodiamond. We show that individual sub-crystallites within a single nanodiamond have distinct optical properties, including spectrally shifted ZPLs as well as differences in CL brightness. The changes between crystallites account for heterogeneity in the SiV$^-$ emission more so than surface structure and grain boundaries within the diamond. In fact, we see little change in emitter brightness at the surface of the nanodiamond compared to the bulk. Finally, we show for every diamond studied that a decrease in CL brightness is associated with a ZPL redshift; this effect occurs in nanoscale spatial locations within a single particle, and we correlate this effect with a 2-5\% lattice contraction by mapping strain at the nanoscale. We propose that this lattice strain is likely caused by defect density changes, and the defect induced strain subsequently shifts the ZPL energy.

\subsection{Grain Boundaries support unstable various SiV$^-$ related emission centers}

\begin{figure}[H]
\begin{center}
\includegraphics[width=90mm]{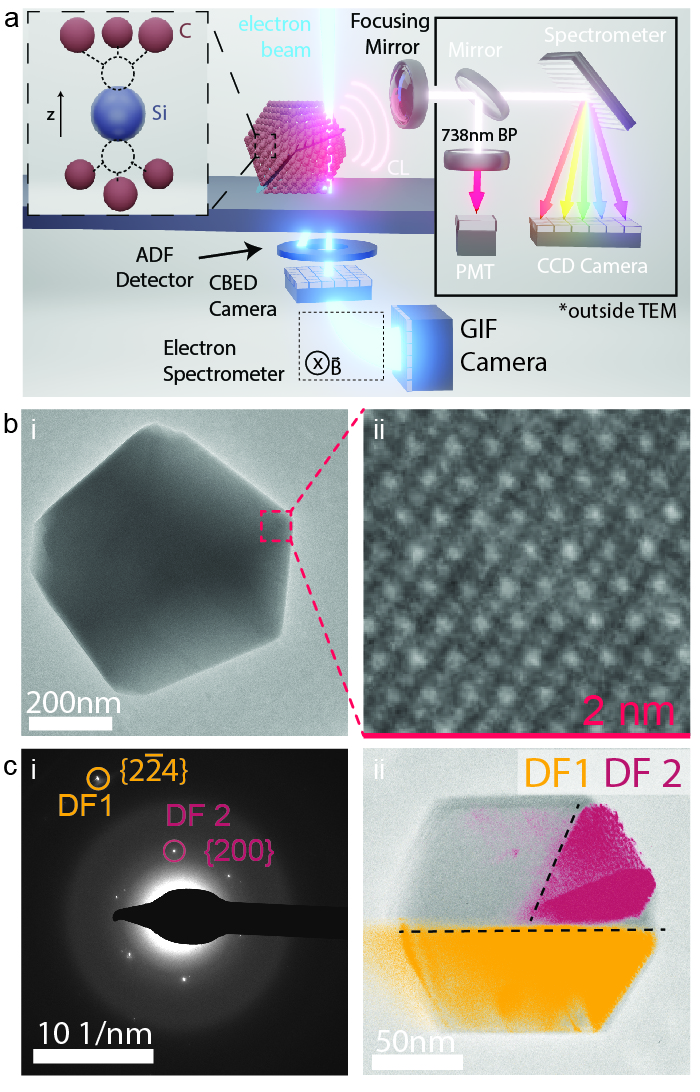}
\end{center}
\caption{ \textbf{CL spectroscopy of high quality multi-crystalline nanodiamonds}. (a) STEM CL Schematic (b) i-ii: TEM of typical Nanodiamond containing SiV$^-$s ii: HRTEM of diamond surface (c) i: shows a single particle Selected Area Diffraction Pattern (SAED) indicating the particle’s multi-crystallinity. ii: We perform off-axis objective aperture dark field imaging and overlay two dark field images on the bright field TEM image. Dashed lines are used as a guide to the eye to indicate roughly the underlying grain boundaries within the nanodiamond.}
\label{fig:Schematics}
\end{figure}

Figure \ref{fig:Schematics}a shows a schematic of the experimental setup. A condensed STEM beam penetrates the nanodiamond and is analyzed \textit{via} multiple EM methods such as: Annular Dark Field (ADF) imaging, Convergent Beam Electron Diffraction (CBED) patterns, and EELS spectra. Excited carriers generated by the STEM beam (mostly a result of decaying bulk plasmons at 34 eV~\cite{meuret_photon_2015-1}) populate the SiV$^-$ defect (structure inset in Figure \ref{fig:Schematics}a)  excited state leading to optical CL emission, that we analyze with grating-based spectroscocpy outside of the TEM column. We study nanodiamonds (see Methods) that are grown by CVD synthesis providing highly crystalline domains as shown in Figure \ref{fig:Schematics}b. The high crystal quality throughout the particle is further evident from the HRTEM image of the nanodiamond close to the surface shown in Figure \ref{fig:Schematics}bii. We confirm the multi-crystallinity of our nanodiamonds through Selected Area Diffraction Pattern (SAED) analysis (Figure \ref{fig:Schematics}c). We note that we studied various sizes of nanodiamonds, including large scale micron-sized multi-crystalline nanodiamonds and sub-50nm diamonds (see Figure \ref{SI:smalldiamonds}). 

\begin{figure}[H]
\begin{center}
\includegraphics[width=90mm]{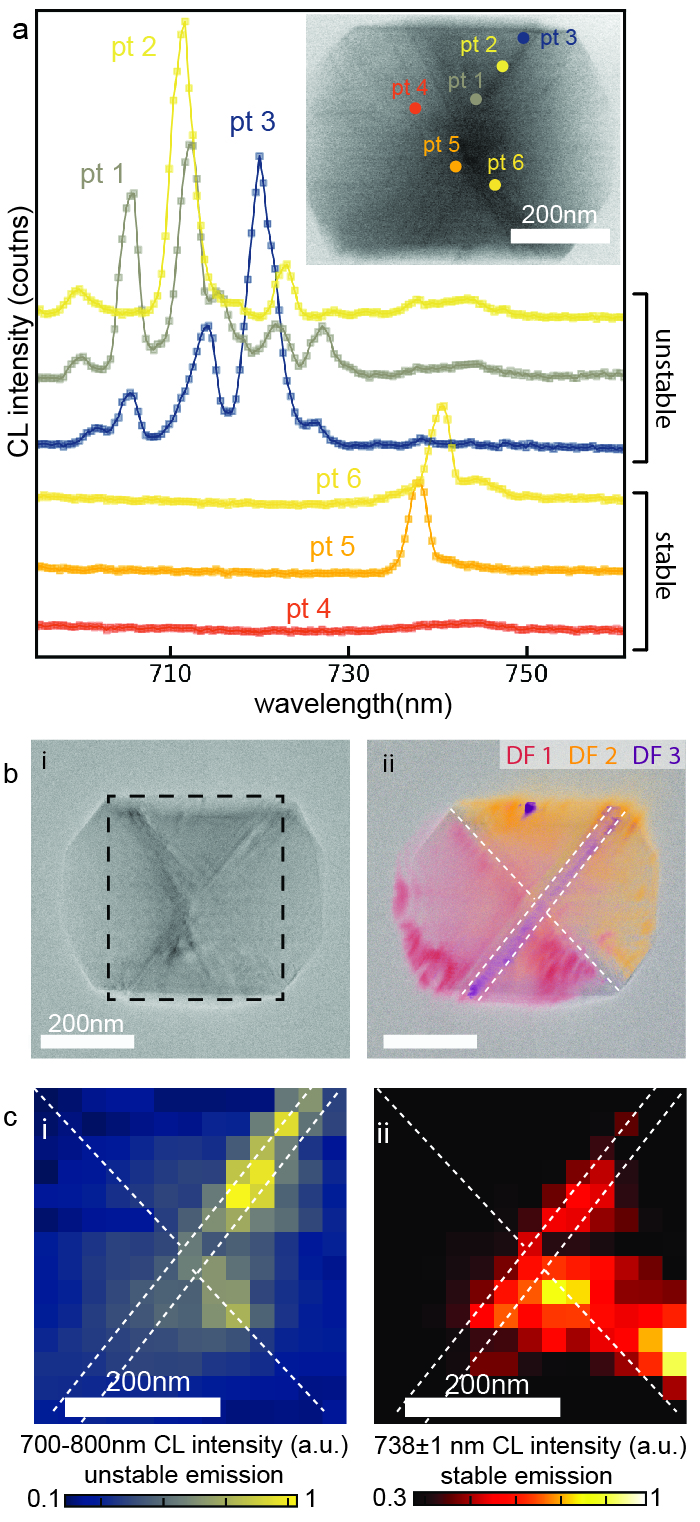}
\end{center}
\caption{ \textbf{Unstable vs stable silicon dopant emission correlates with grain boundaries} (a) CL point spectra taken when a STEM probe at positions pt. 1-6, corresponding to the points in the inset BF STEM image, sale bar 200 nm. (b) i TEM image of nanodiamond in figure 2, ii three dark field images combined into one image, overlaid on the TEM image (c) i: 2D map of CL counts for CL wavelengths 700-800 nm, and ii: for CL wavelengths 738$\pm$1 nm, corresponding in space to the black dashed box in (b)i, scale bars 200 nm}
\label{fig:Nonstable}
\end{figure}

We study the heterogeneity observed with SiV$^-$ emission, by taking STEM-CL point spectra at different positions within a large nanodiamond as shown in Figure \ref{fig:Nonstable}. We note that these acquisitions are well below the optical diffraction limit, meaning the spatial resolution provided by STEM-CL is unobtainable with traditional optical spectroscopic mapping. Additionally, the energy resolution needed to observe changes in the SiV$^-$ electronic structure (0.1 meV) is impossible to obtain with even the most advanced monochromated EELS machines. We observe two main classes of emitters inside the same single nanodiamond. First, spectrally differing emission (\textit{e.g.}, the sharp peaks at pt. 1, pt. 2, and pt. 3)  are observed in large, multicrystalline nanodiamonds. These largely variant emitters are previously found to be correlated with SiV$^-$ centers~\cite{ lindner_strongly_2018}  and are indirectly shown to involve imperfect diamond lattices~\cite{bray_localization_2016}; however, their exact structure remains unknown. As shown previously, these non-738 nm emitters are unstable~\cite{ lindner_strongly_2018} and can only be detected within the first few seconds of electron beam irradiation. The emission bleaching is irreversible; even after a week at room temperature, the emitters did not reappear (Figure \ref{SI:Transient}). Second, we identify the split SiV$^-$ divacancy (\textit{$D_{3d}$} symmetry) stable emission, characteristic at 738 nm (pt. 4, pt. 5, and pt. 6). CL brightness, spectral position, and acoustical phonon side bands vary for different point spectra suggesting sub-diffraction limited heterogeneity.

We correlate the CL intensity with the occurrence of grain boundaries. Figure \ref{fig:Nonstable}b shows a TEM image of the nanodiamond in Figure \ref{fig:Nonstable} overlaid with three dark field images to elucidate the particle grain structure. Grain boundaries occur where two colors meet in these combined images, and white dashed lines serve as a guide to the eye. The CL maps taken in the area of the black box for the unstable and stable emitters (c) reveal that i)  the non-738nm emitters reside near or within grain boundaries and ii) pockets of intense SiV$^-$ emission occur only over a range of a 100 nm, and this intensity correlates with the grain boundaries within the multicrystalline particle. See Figure \ref{SI:NNMF} for further data on large nanodiamonds. The unstable signal residing near or within grain boundaries has been hypothesized previously~\cite{bray_localization_2016}. The observed heterogeneity possibly indicates that silicon incorporation during the CVD growth varies and may be aided by 2D crystal defects; this supports theoretical findings that defect formation at grain boundaries is more energetically favorable~\cite{zapol_tight-binding_2001, lebedev_microstructural_2020}. 


\subsection{Sub-crystallites exhibit distinct optical properties}

\begin{figure}[H]
\begin{center}
\includegraphics[width=110mm]{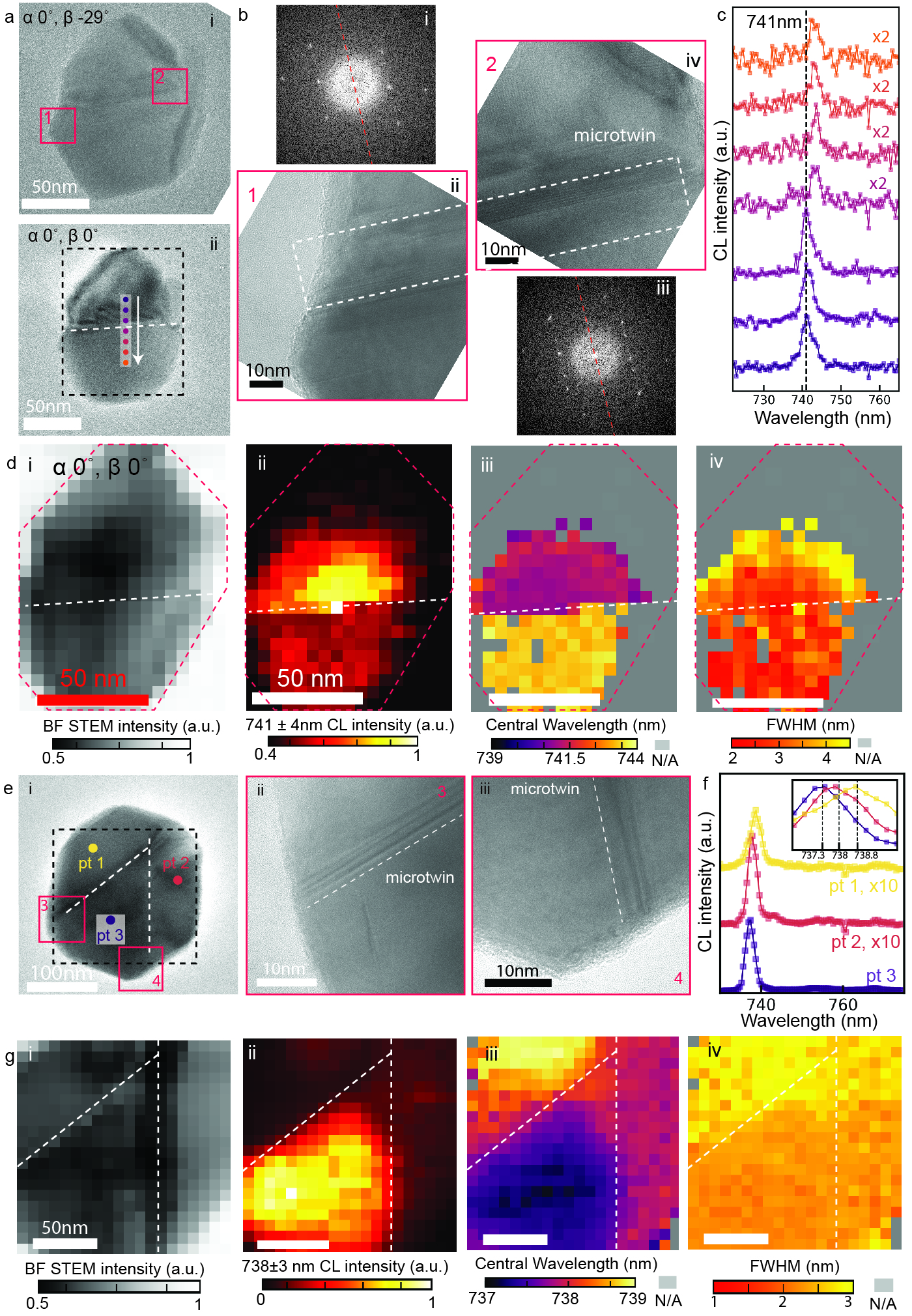}
\end{center}
\caption{ \textbf{Sub-crystallites exhibit distinct optical properties at deep subwavelength spatial volumes} (a) i: Bright field TEM image with 0\degree\ alpha, 29\degree\ beta goniometer stage tilts  ii: Bright field TEM image with 0 alpha, 0 beta goniometer stage tilts, scalebars 50 nm (b) i and iii: FFTs of  the cropped TEMs in b ii and b iv,  corresponding to red squares 1 and 2 shown in ai. (c) point spectra with the electron probe at circles in aii (corresponding in color)  (d) i: 2D map of BF STEM counts ii: 2D map of CL counts for CL wavelengths 741$\pm$4 nm, iii: 2D map of central wavelength of SiV$^-$ emission, iv: 2D map of FWHM of SiV$^-$ emission. 2D maps correspond to black dashed box in aii. Scale bars 50 nm. (e) i: TEM image. ii-iii: cropped TEM images corresponding to red boxes in ei. (f) point spectra take at circles in ei (corresponding in color), inset is a zoom in on ZPLs of point spectra. (g) i: 2D map of BF STEM counts ii: 2D map of CL counts for CL wavelengths 738$\pm$3 nm, iii: 2D map of central wavelength of SiV$^-$ emission iv: 2D map of FWHM of SiV$^-$ emission. 2D maps correspond to black dashed box in (e)i.}
\label{fig:Subcrystallites}
\end{figure}

We focus on the observed spectral shifts in the SiV$^-$ defect emission by collecting 3D hyperspectral CL maps at higher resolution (5.4 nm) from particles with well-defined grain structure. From the HRTEMs shown in \ref{fig:Subcrystallites}b, we can see that a grain boundary runs across the entire nanoparticle. Specifically, in \ref{fig:Subcrystallites}bii and biv, we see a striped pattern cutting the particle in half that is indicative of a microtwinned grainboundary (delineated by the white dashed box)~\cite{shechtman1993growth,shechtman1994twin,butler2008mechanism}. Fast Fourier Transforms (FFTs) of images 1 and 2 confirm the particle contains a twin boundary (bi and biv)~\cite{shechtman1993growth}. Finally, FFTs of the cropped red and blue square regions in images 1 and 2, produce single crystal \{110\} patterns (see Figure \ref{SI:p10structure} for analysis), where the angle of rotation between these two patterns is measured to be 70.9$\pm$0.5\degree\ (Figure \ref{SI:matti_twin}), confirming the microtwinned boundary is a low energy $\Sigma$3. The $\Sigma$3 twin boundary preserves the tetrahedral units both in direction and bond length of the diamond structure, and therefore the crystal structure remains coherent~\cite{shechtman1993high}.

We observe discrete spectral heterogeneity in CL emission depending on which subcrystallite we probe. Figure \ref{fig:Subcrystallites}c reveals that point spectra display red-shifted SiV$^-$ emission, with a 2 nm wavelength shift across the boundary, concurrent with a reduction in CL intensity by a factor of two. We extract spatial maps of the total CL intensity at 738nm, the central wavelength, and the FWHM of the ZPL from Lorentzian fits to the 3D hyperspectral maps. The results are shown in Figure \ref{fig:Subcrystallites}di-iv, together with the Bright Field STEM intensity. All maps cover the area of the black dashed rectangle in aii. Figure \ref{fig:Subcrystallites} dii confirms that the twinned boundary separates two regions of distinct optical properties with a discrete reduction in the SiV$^-$ emission intensity by about 40\% from the top to the bottom crystallite. Remarkably, this drop is consistent across the entire length of the boundary and occurs within 1 pixel length (5.4 nm) across the boundary. We note that the reported excited charge carrier diffusion lengths in nanodiamond CL experiments exceed 5nm~\cite{kozak2012large, tizei_spectrally_2012}, suggesting that the microtwinned grain boundary acts as a barrier to carrier diffusion.

Similar to the CL intensity, we see a sharp, discrete change in central wavelength across the particle grain boundary (diii), where the top crystallite's emission is centered around 741.3 nm, and the bottom crystallite's emission is centered around 743.3 nm. These changes in ZPL energy are, however, not correlated with any significant change in the FWHM of the 738 nm emission, suggesting that the mechanism inducing red-shifting causes no emission linewidth broadening ( Figure \ref{fig:Subcrystallites} div).
 
To illustrate how inhomogeneity is caused by sub-crystallites, we include similar data from another nanodiamond (Figure \ref{fig:Subcrystallites}ei). This diamond has two microtwinned grain boundaries instead of one. These boundaries are shown in the HRTEM images in \ref{fig:Subcrystallites}eii-iii, where the HRTEMs correspond in space to the red squares in ei. Now, if we position a STEM probe at points 1,2 and 3 (denoted in ei and corresponding in color), we see that three distinct, spectrally shifted emission profiles are produced (pt. 1 at 738.8 nm, pt. 2 at 738 nm, and pt. 3 at 737.3 nm). The 3D hyperspectral mapping is shown in gi-iv; the approximate crystal structure is indicated by the white dashed lines, which separate the nanodiamond into three separate crystallites, each spectrally shifted from the other. Furthermore, we again observe that a majority of CL emission is being produced by one crystallite (in this case the bottom left crystallite) compared to the rest of the particle. These results are consistent across multiple particles, and the rest of the data can be found in Figures \ref{SI:p12}-\ref{SI:p4}. We conclude that the differences between individual crystallite domains is a dominant contributor to inhomogeneous broadening in SiV$^-$ defect ensembles in nanodiamonds; this is further confirmed by the homogeneity of SiV$^-$ emission found in small diamonds (see Figure \ref{SI:smalldiamonds}).

\subsection{Emitter CL intensity variations at the nanoscale}

\begin{figure}[H]
\begin{center}
\includegraphics[width=165mm]{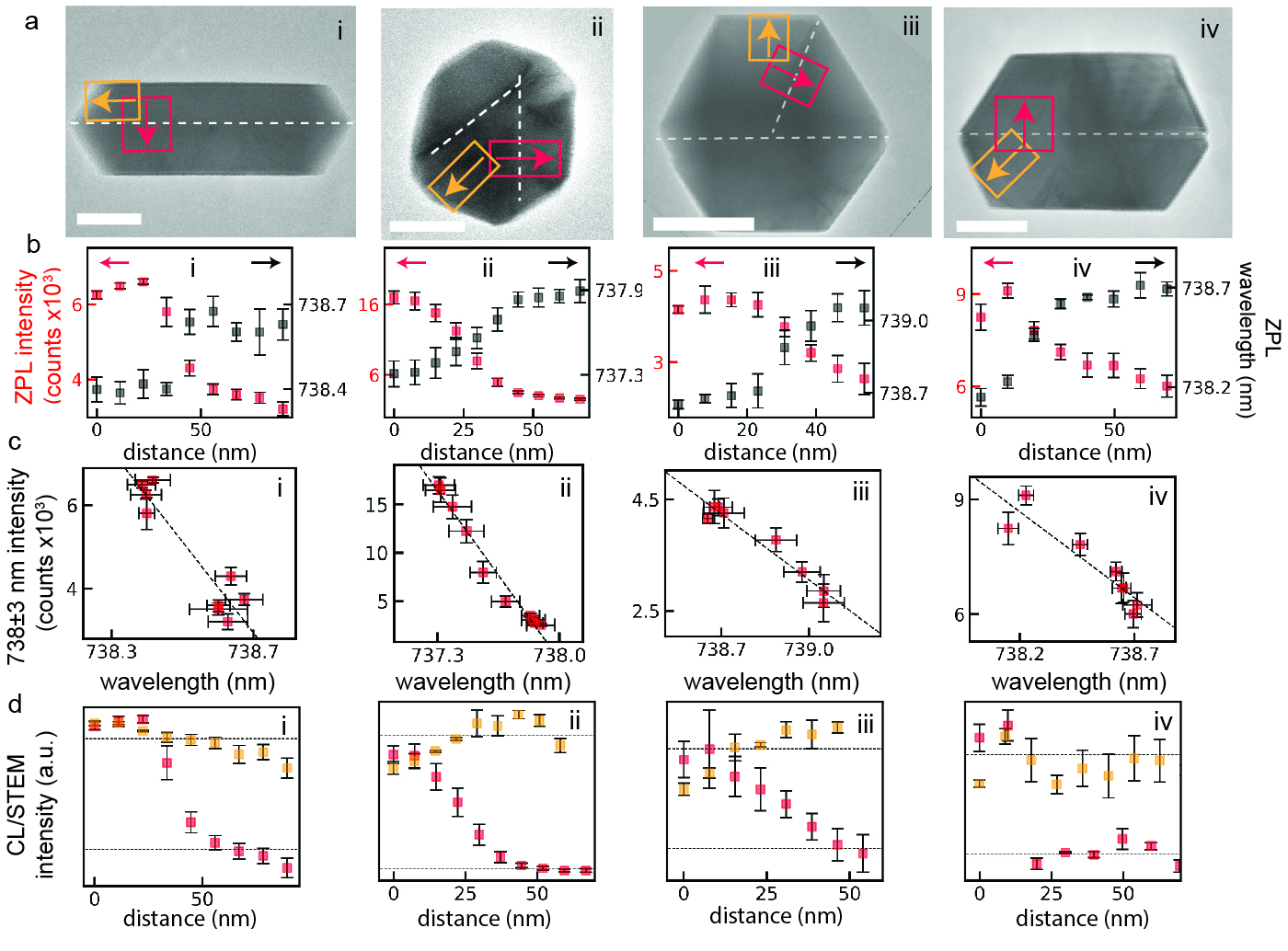}
\end{center}
\caption{ \textbf{ZPL energy and brightness correlations reveal multiple findings} (a) i-iv four example nanodiamonds with prominent grainboundaries delineated by white dashed lines (b) i-iv profile along red arrow of both the 738$\pm$3 nm CL counts in red (left axis) as well as ZPL central wavelength in black (right axis). (c) i-iv scatter plots of 738$\pm$3 nm CL counts vs ZPL central wavelength at each pixel of the hyperspectral map pixels in the red box . (d) i-iv profiles along the red and orange yellow arrows (corresponding in color), of the ratio of 738$\pm$3 nm CL counts to that of the STEM camera intensity. Bins are created parallel to the boundary (perpendicular to red/orange arrow), and error bars represent a standard deviation of the bin. Scale bars 100 nm.}
\label{fig:lineprofiles}
\end{figure}

 We analyze the ZPL energy and brightness along linescans ( \ref{fig:lineprofiles} ai-iv) for four representative nanodiamonds. We confirm the abrupt changes in the CL intensity and ZPL energy discussed above in the linescans along the red arrows (bi-iv). Remarkably, for most particles, the intensity and ZPL energy are linearly correlated as shown in ci-iv. Additional nanodiamonds, including those with differing behavior are explored in the Figure \ref{SI:Lineplots}. The consistent trend between emitter brightness and ZPL wavelength suggests that the same crystal perturbation is affecting both the energy and the brightness of the SiV$^-$ defect. Changes in CL emission intensity can be due to a number of factors including: excitation efficiency, thickness, electron-phonon coupling, crystal orientation, local dielectric environment, defect density, nonradiative relaxation pathways, or defect charge state blinking~\cite{mahfoud_cathodoluminescence_2013,brown1999transmission, monticone2013systematic, zhang2014silicon, choy2011enhanced, collins1990spectroscopic,castelletto2011radiative}. Using a combination of EELS, CL, and electron diffraction, we explore and account for sources of brightness heterogeneity in Figures \ref{SI:EELS}-\ref{SI:LAMMPS}, ensuring that the change in brightness is due to either a modification of the defect's quantum yield (QY), or a change in defect density. 

To account for the different efficiencies in excitation of bulk plasmons owing to spatially varying thickness, we normalize the CL intensity by the STEM camera counts as shown in Figure \ref{fig:Subcrystallites}di-iv. This normalization serves as an effective particle thickness control~\cite{brown1999transmission}. The normalized data along the red path (across the grain boundary) suggests that the changes in CL brightness between different boundaries can be ascribed to local differences in incorporation density or intrinsic QY of the SiVs. The corresponding data along the orange path (across the same boundary towards the surface) shows no significant changes, demonstrating the absence of surface quenching or changes in defect density, even for the pixel closest to the surface. Indeed, we see similar spatially-invariant brightness in small nanodiamonds ($<$50 nm), where a step size of 2.9 nm is used between pixels, ensuring that our probe interacts with the first few nm of nanodiamond material (Figure \ref{SI:smalldiamonds}). The same trends of minimal surface quenching and grain-dependent brightness are seen in a majority of particles studied and can be found in Figure \ref{SI:Lineplots}. Shallow defect incorporation and the defect’s subsequent interaction with the surface has been the focus of recent studies, where it has been shown that surface treatment can improve SiV$^-$ optical properties, such as narrower linewidths and brighter emission~\cite{lang_long_2020}. Shallow defects are important for device deployment, due to their optical addressability and enhanced coupling to nanophotonic structures~\cite{sipahigil2016integrated,lang_long_2020}. Our analysis shows, that even surface-near SiVs (2.9nm) exhibit unchanged relative CL brightness compared to the bulk.

\subsection{Nanoscale strain mapping}

We now focus on the structural origin of the observed spatially-variant CL properties. We employ 4D STEM analysis on multiple particles to deduce if the strain states of individual sub-crystallites are responsible for the CL changes. 4D STEM datasets can be used to map lattice strain at nanometer length-scales by measuring changes in Bragg disk positions across multiple Convergent Beam Electron Diffraction (CBED) patterns, with strain resolutions down to 6x$10^{-4}$~\cite{beche_strain_2013, beche2009improved, armigliato2006convergent}. We show the analysis for a nanodiamond with a large boundary running across its center.

\begin{figure}[H]
\begin{center}

\includegraphics[width=150mm]{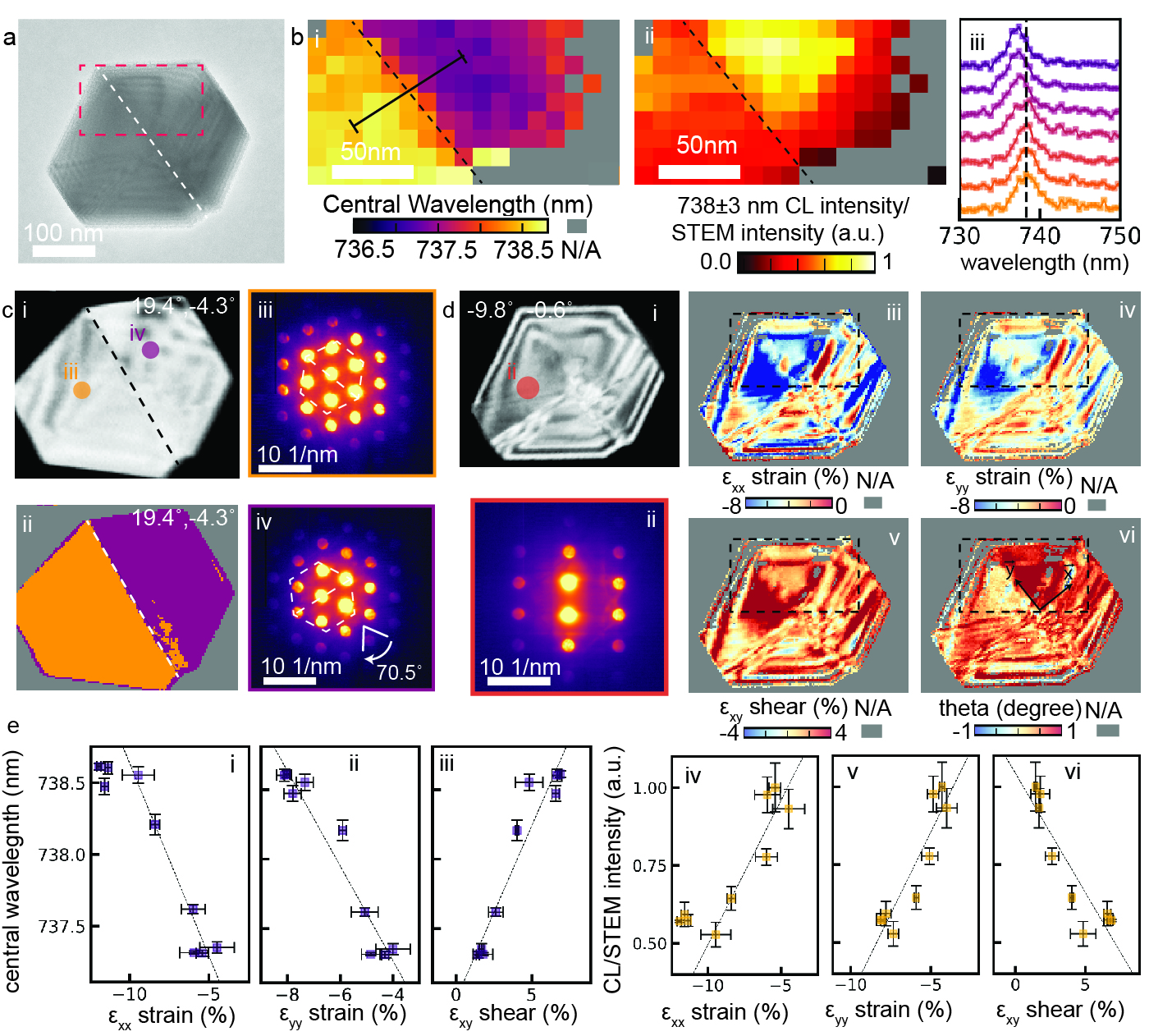}

\end{center}
\caption{ \textbf{SiV$^-$ optical properties correlate with strain  at the nanoscale} (a)  TEM image (b) 2D hyperspectral maps taken at black dashed box in a, i Central wavelength of lorentz fit, ii: 738$\pm$3 nm summed intensity, iii: CL point spectra taken along black line in bi. (c) 4D STEM data taken  at goniometer stage tilts 19.4\degree\, -4.3\degree\ i: Virtual DF image, ii: Twin boundary crystal classification, iii-iv: CBED patterns from spots in i, corresponding in color. (d) 4D STEM data taken  at goniometer stage tilts -9.8\degree\, 0.6\degree\ i: Virtual DF STEM image, ii: CBED pattern take at spot in i, iii-vi: $\epsilon_{xx}$ strain, $\epsilon_{yy}$ strain, $\epsilon_{xy}$ shear, rotation. (e) i-iii: ZPL wavelength vs $\epsilon_{xx}$ strain, $\epsilon_{yy}$ strain,  and $\epsilon_{xy}$ shear. iii-iv: ZPL intensity normalized by STEM counts vs $\epsilon_{xx}$ strain, $\epsilon_{yy}$ strain,  and $\epsilon_{xy}$ shear, for CL data taken along black line in bi that crosses the middle boundary of the particle.}
\label{fig:Strain}
\end{figure}

In  Figure \ref{fig:Strain} bi-ii, we show the 738$\pm$3 nm CL counts, and the ZPL central wavelength, respectively, corresponding in space to the dashed  red box in a. As observed in previous particles, we can clearly see that the crystal domain boundary indicated by the black dashed line separates the particle's optical properties, both in brightness and in ZPL energy.

Multiple 4D STEM data sets were taken of the entire particle; a virtual Dark Field STEM image is produced from this data set and shown in (ci), which delineates where strain can be analyzed. We orient the particle along the ${110}$ zone axis, and can clearly observe that the central boundary is a $\Sigma$3 twin. We classify each CBED pattern in the dataset to produce an image of the two grains and their boundary, shown in (cii) as orange and maroon, where the corresponding representative CBED patterns are shown in ciii-civ. Because the crystals are rotated by 70.5\degree\ with respect to each other along the $\{110\}$ zone, we re-orient the particle along $\{211\}$ to perform comparative strain analysis of the entire particle. We perform strain analysis of the $\epsilon_{xx}$ and $\epsilon_{yy}$ strain, the $\epsilon_{xy}$ shear,  and $\theta$ rotation, show in \ref{fig:Strain} (diii-vi). The middle twin boundary separates the crystallite into distinctly strained regions, where the  top right crystallite is expanded and rotated compared to the bottom left (Figure \ref{fig:Strain} diii-vi specifically).

STEM-CL enables the correlation of the ZPL intensity and energy with the local strain of the nanodiamond, allowing us to draw two trends. First, we identify a blue shift of the ZPL and an increase in CL intensity, with a positive $\epsilon_{xx}$ and $\epsilon_{yy}$ strain (\textit{i.e.} a lattice expansion) across the middle boundary (Figure \ref{fig:Strain} e i-ii and iv-v). This trend across domain boundaries is consistent across multiple particles (shown in \ref{SI:p8Strain} and \ref{SI:p17_new_Strain}). Second, with a positive change in  $\epsilon_{xy}$ shear we observe a red shifting ZPL, and a decreasing intensity of the SiV (Figure \ref{fig:Strain} e iii and vi) and is again consistent in multiple particles (shown in \ref{SI:p8Strain}). The $\overrightarrow{x},\overrightarrow{y}$ basis is oriented along the twin boundary, as shown in Figure \ref{fig:Strain} dvi.

The consistent trend we observe across domain boundaries is likely a change in the dopant incorporation rate of the different crystallites. Defect incorporation (specifically nitrogen and boron defects) into diamond  can be facet-dependent~\cite{wilson_impact_2006,szunerits_raman_nodate,burns_growth-sector_1990}. Intuitively, a large change in the defect density within a crystal lattice will lead to large changes in the lattice's stress state~\cite{anthony_stresses_1995, rossi_micro-raman_1998}. In a diamond lattice, changes in lattice stress states have been experimentally measured across grain boundaries \textit{via} Raman spectroscopy of the optical phonon~\cite{rossi_micro-raman_1998,vlasov_analysis_1997}, where a grain boundary has been shown to separate regions of distinct Raman frequencies. Silicon defects in diamond expand the lattice due to Si’s larger diameter, which we confirm using molecular dynamics simulations (figure \ref{SI:LAMMPS})~\cite {erhart_analytical_2005}. Such a mechanism could explain the trend observed across domain boundaries, where lattice contraction produced less intense CL, due to a smaller SiV$^-$ defect density. This mechanism would explain why we observe shifts in intensity and energy, even across a $\Sigma$3 twin boundary (such as the one identified in Figure \ref{fig:Subcrystallites}, Figure \ref{fig:Strain}, or in Figure \ref{SI:p3newStrain}). An ideal $\Sigma$3 remains coherent, and therefore should induce little lattice distortion, which suggests that varying defect incorporation rates across the $\Sigma$3 boundary is the most plausible source of strain in the particle. Interestingly, if this mechanism is correct, our results produce opposite shifts of the ZPL energy compared to Meesala et. al.\cite{meesala_strain_2018}, possibly indicating defect induced stress affects SiV$^-$ emission differently than mechanically-induced stress from an AFM tip.

We also observe some nanodiamonds with CL ZPL strain correlations \textit{within} single crystallites of nanodiamonds, which have the opposite correlation to those found \textit{across} grain boundaries (see Figures \ref{SI:p8Strain} and \ref{SI:p3newStrain}). The trends we observe within a crystallite could be a result of changes of the QY of the SiV$^-$ defect itself.  The QY of the SiV$^-$ defect is estimated to be 5\%~\cite{turukhin_picosecond_1996}, indicating that a vast majority of energy is not radiated into the ZPL channel. Although the SiV$^-$ QY is consistently low, the values of QY can vary between individual emitters~\cite{neu_photophysics_2012}, which has often been attributed to defect implantation crystal defects, imperfections at the crystal surface, bulk structural defects, or nondiamond phases and graphitic bonding~\cite{smith2010effects,grudinkin2012luminescent,lang_long_2020}. Notably, our results exclude these factors as the main contributors to CL intensity changes, as we see no decreases in CL emission approaching the grain boundaries and surfaces, and no change in diamond bonding across or within crystallites (see Figure \ref{SI:EELS_boundary1}-\ref{SI:EELS_boundary3}). Exploring nanoscale QY changes within a single crystal lattice is the subject of future work.

In summary, our results show that grain boundaries within nanodiamonds promote largely heterogeneous emission from defects associated with silicon dopants in diamond. Individual subcrystallites within a single nanodiamond have differing SiV$^-$ optical properties with up to 2nm spectral shifts and 70\% brightness changes, and are likely the largest contributor to inhomogeneous broadening of SiV$^-$ ensembles. Changes found between sub-crystallites largely overshadow any changes due to surface proximity. We find that in a large majority of particles, ZPL energy and intensity are positively correlated and are spatially correlated with large, static strains within the diamond lattice, that permanently shift the ZPL of the emission, without altering the electron-phonon coupling or homogeneous broadening of the emission. The ZPL intensity can change by 50\% or more, and we propose this change occurs due to multiple mechanisms, including the presence of defect density gradients within single nanodiamonds, and possible changes in the defect’s emission pathways. These findings elucidate the structural sources of heterogeneity of SiV$^-$ optical emission, and can inform materials design and synthesis of future quantum sources and sensors.

\section{Methods}

Nanodiamonds were synthesized via plasma enhanced chemical vapor deposition (CVD), and grown directly on the amorphous SiO2 20 nm TEM grids (TEM windows). CL data was taken with an FEI Titan aberration corrected environmental transmission electron microscope (TEM), using the Gatan Vulcan Cathodoluminescence in-Situ holder, at 80kV of accelerating voltage. All CL data sets were taken at a temperature of 100 K. A majority of the 3D hyperspectral CL maps were taken with 100pA of current, and 5 seconds of dwell time (see Supplementary Info for exact parameters), with a 14 mrad convergence angle. Dark field images were taken at 80kV, with both on axis and off axis objective aperture configurations. 4D STEM data sets were acquired with Gatan's Oneview camera at 300kV and room temperature with a current of 30pA, a camera length of 480mm, and a convergence angle of 1.8 mrad. Data analysis was performed in python, utilizing multiple common packages, such as numpy, scipy, and matplotlib. Non-negative matrix factorization and general data loading and viewing was done with hyperspy~\cite{francisco_de_la_pena_2016_58841}. 4D STEM analysis, including virtual dark field images as well as local strain mapping was performed with py4dstem~\cite{savitzky2021py4dstem}.

\begin{acknowledgement}
We appreciate thoughtful feedback from Debangshu Mukherjee, Parivash Moradifar, Briley Bourgeois, as well as Chris Ciccarino on the manuscript. All authors gratefully acknowledge the National Science Foundation for funding. D. A. was supported through the NSF GRFP, with experiments supported by the DOE Q-NEXT Center. Diamond synthesis by S.L. and N.M. was supported by a SLAC Laboratory Directed Research and Development (LDRD) program. The research at SIMES is supported by the DOE office of basic energy sciences, Division of Materials Science and Engineering. Part of this work was performed at the Stanford Nano Shared Facilities (SNSF), supported by the National Science Foundation under award ECCS-2026822. H.U.  was supported in part by the 'Photonics at Thermodynamic Limits' Energy Frontier Research Center funded by the U.S. Department of Energy, Office of Science, Office of Basic Energy Sciences under Award Number DE-SC0019140. J.A.D. also acknowledges salary support from the DOE Q-NEXT Center.

\subsubsection{Author Contribution:}
D.A., J.A.D. S.L. and N.M. conceived the idea. S.L. grew the nanodiamonds and D.A. designed and executed all electron microscopy experiments. Y.L. assisted with dark field TEM. M.T. assisted with EELS analysis and J.D., R.C. and Z.X.S. assisted with diamond growth. R.S. facilitated TEM interpretation. D.A., H.U., S.L., and J.A.D. interpreted the data and wrote the initial draft, and J.A.D. supervised the project. All authors contributed to the final manuscript.

\subsubsection{Competing interests:} Authors declare no competing interests. 
\end{acknowledgement}


\newpage

\newpage

\newpage

\newpage


\newpage


\newpage
\bibliography{Main.bib}

\makeatletter\@input{xx.tex}\makeatother
\makeatletter\@input{output.tex}\makeatother

\end{document}


\flushbottom

\newpage
\subsection{Various ZPL SiV related emitters are unstable under the electron beam}

\begin{figure}[!hb]
\centering
\includegraphics[width=130mm]{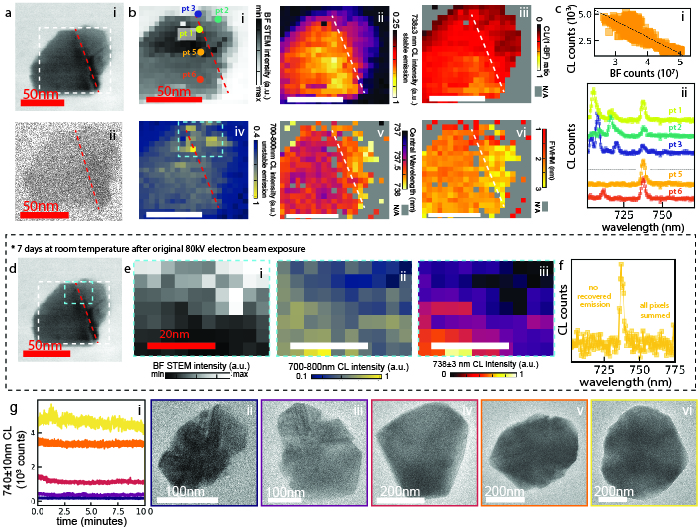}
\caption{ (a) i STEM BF image, ii TEM image.  (b) i-vi BF intensity, 738$\pm$3 nm intensity, 738$\pm$3 nm intensity normalized by BF intensity, 700-800nm CL counts, ZPL wavelength from lorenztian fit, FWHM from lorenztian fit, hyperspectral data corresponding in space to the white dashed box in ai. (c) i 738$\pm$3 nm counts plotted against BF STEM counts, ii point spectra taken at spatial locations show in bi. (d) After one week, STEM BF image, (e) i-iii hyperspectral data corresponding to cyan box in (d) and biv,  BF intensity, 700-800nm CL counts,  738$\pm$3 nm intensity. (f) all pixel spectra summed from cyan box. (g) 740 $\pm$ 10 nm CL counts vs time for the 5 particles shown in ii-vi, TEM illumination, corresponding in color to the TEM images shown in  ii-vi.}
\label{SI:Transient}
\end{figure}

As stated in the main text, the varying central wavelength emission centers (between 700 nm and 730 nm, ususally) observed in these particles are unstable in nature. These emitters were previously shown to be prone to charging/blinking~\cite{lindner_strongly_2018}. In Figure \ref{SI:Transient}, we demonstrate how these emitters are extremely beam unstable. In Figure \ref{SI:Transient} (c)ii, we show multiple spectra (pts1,2 and 3) that exhibit emitters at 705, 707, 716, and 720nm. These emitters are spatially located along the grain boundary indicated by the red dashed line in \ref{SI:Transient} (a)i. After the initial exposure of 5 seconds/pixel at 100pA beam current, we left the sample for 1 week at room temperature. If these emitters were simply experiencing charging/blinking, a week at room temperature should be enough time to recover emission~\cite{lindner_strongly_2018}. Since we recover no emission, as shown in (f) where we sum all pixels of data, we conclude that the beam exposure completely destroyed the emitters. In contrast to this instability, emission centered at 738 nm remains fairly constant over minutes of beam exposure, as shown in Figure \ref{SI:Transient} (g)i for particles shown in (g)ii-vi. Here, emission at 738 nm is captured continuously for 10minutes, where we see very little decay in the intensity of the signal over this time period, for multiple particles, differing in size. This indicates that the 700-730 nm emitters are structurally unstable compared to the D$_{3d}$ SiV$^{-}$ defect emitter at 738 nm.

\clearpage

\subsection{Large Diamond NMF analysis}

\begin{figure}[!hb]
\centering
\includegraphics[width=130mm]{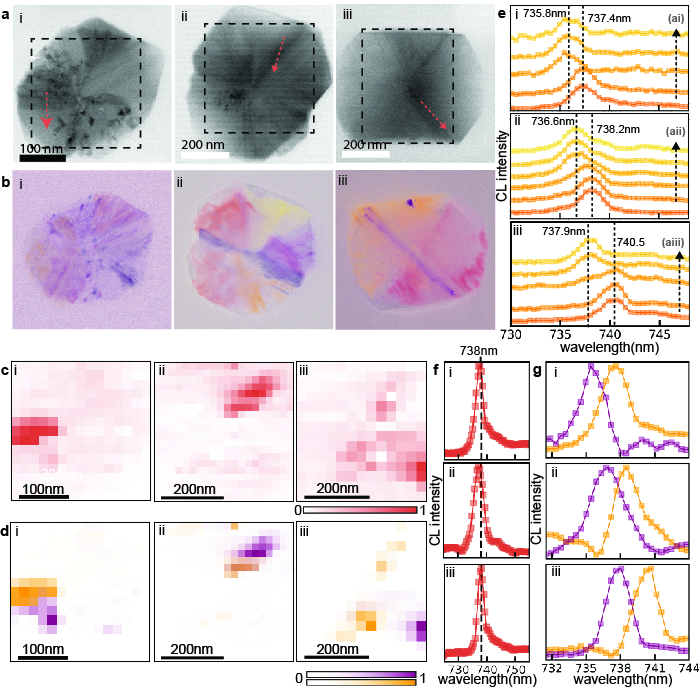}
\caption{(a)i-iii TEM images of 3 nanodiamonds,(b) i-iii DF images overlaid on TEM images (c) i-iii NNMF maps corresponding to (f)i-iii for particles i-iii. (d) NNMF maps for a two component system corresponding to in color to basises shown in (g)i-iii for particles i-iii.(e) point spectra taken along the red dashed arrows shown in (a)i-iii}
\label{SI:NNMF}
\end{figure}

The spectral change between pts 2 and 3 in Figure 2 were explored with non-negative matrix factorization (NMF) to decompose the 3D data set~\cite{hayee_revealing_2020}. In figure \ref{SI:NNMF} we perform NMF data decomposition on 3 large nanodiamonds, but will focus here on particle (a)iii, because we analyze this diamond in Figure 2. In figure \ref{SI:NNMF} (c)iii we allow the NMF algorithm 1 component to approximate the 738 nm defect emission. Figure\ref{SI:NNMF} (c)iii and \ref{SI:NNMF} (f)iii show the NMF weight map and the NMF basis spectra, respectively. However, if we allow the NMF algorithm to use 2 components to factorize this data instead of one, the algorithm naturally splits the basis spectrum into two spectrally resolved emissions profiles (purple and gold), shown in \ref{SI:NNMF}(g)iii. In (d)iii, we plot both weight maps corresponding in color to the spectra in (g)iii. The two spectra are shifted by ~3nm spectrally from each other, even though they reside in the same nanodiamond. Interestingly, the two different emission profiles are grouped spatially within the particle, and again correlate with the underlying grain boundaries. Note that these weight maps (gold and purple), when combined and normalized, remake the weight map shown in (c)iii. Interestingly, it is noticeable that the two observed 738 nm SiV$^{-}$ emitters found in each of these three nanodiamonds differ slightly from each other. This indicates that in addition to subgrain variations within a single particle, there are absolute differences in emission between each CVD grown nanodiamond.

\clearpage

\clearpage

\subsection{Accounting for all potential factors of CL brightness}

Using EELS, CL, and electron diffraction, we can account for potential sources of CL intensity variations and conclude the variation is a manifestation of a change in the radiative pathways of the SiV$^-$ excited state.

\begin{figure}[!hb]
\centering
\includegraphics[width=130mm]{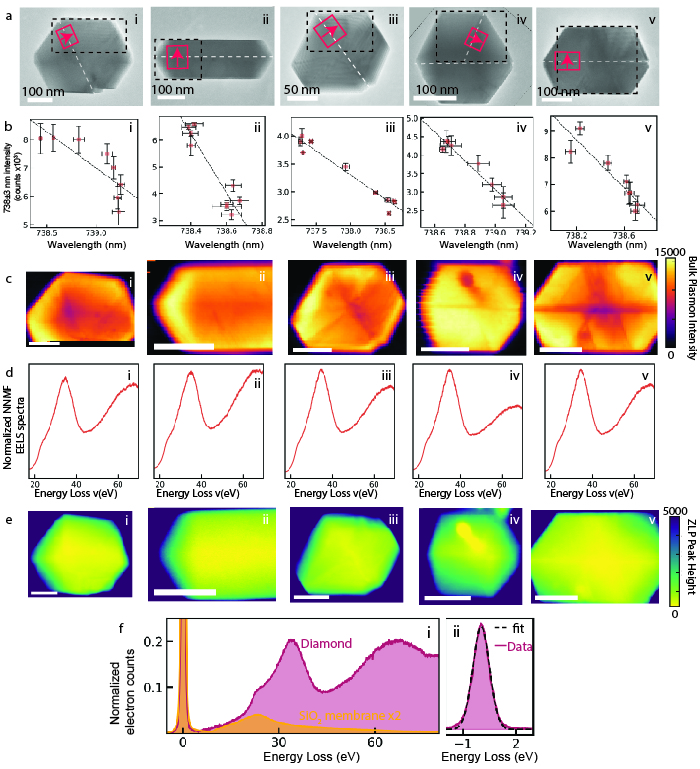}
\caption{(a) i-v TEMs for particles 12-18. (b) i-v 738$\pm$3nm CL intensity counts versus the ZPL central wavelength of a lorentzian fit, as seen in Figures \ref{SI:Lineplots} and \ref{SI:Lineplots2}. (c) i-v Bulk Plasmon maps corresponding to the spectra in (d). (d) i-v Bulk plasmon spectra.  (e) i-v  2D map of Zero Loss Peak gaussian height fits of the EELS spectra (f) i: Example EELS spectra of the substrate and nanodiamond. ii  example of ZLP and gaussian fit used to make maps in (e). scale bars 100nm.}
\label{SI:EELS}
\end{figure}

First, bulk plasmon resonances, the volume plasmon resonance produced at Re$\{\epsilon\}$=0, generate a majority of the excited carriers that produce defect emission~\cite{meuret_photon_2015-1}. This can be easily understood by taking EELS spectra with the probe position penetrating a nanodiamond.  In Fig. \ref{SI:EELS}(f)i, at 34 eV energy loss, we see that there is a maximum in the EELS spectra, meaning that a majority of inelastically scattered electrons impart energy into the bulk plasmon mode. We can quantitatively map the relative number of excited carriers produced in these nanodiamonds by mapping the intensity of the bulk plasmon mode produced in EELS; here we show bulk plasmon maps of 5 particles measured (particles 12, 13, 14, 17,and 18) in Figure \ref{SI:EELS}(c). Across the boundary of the particle, we see no relative change in the excitation of the bulk plasmon mode. Additionally, by monitoring the attenuation of the Zero Loss Peak (ZLP) in EELS, we can also quantitatively map the thickness of diamond the electron probe passes through~\cite{brown1999transmission}. In \ref{SI:EELS}(e)i-v, we show the  Zero Loss Peak attenuation (analogous to the thickness) maps, which don’t show major thickness changes across the boundary. These measurements allow us to rule out that an excitation efficiency change is causing the change in CL counts across the boundary.

\clearpage

\begin{figure}[!hb]
\centering
\includegraphics[width=140mm]{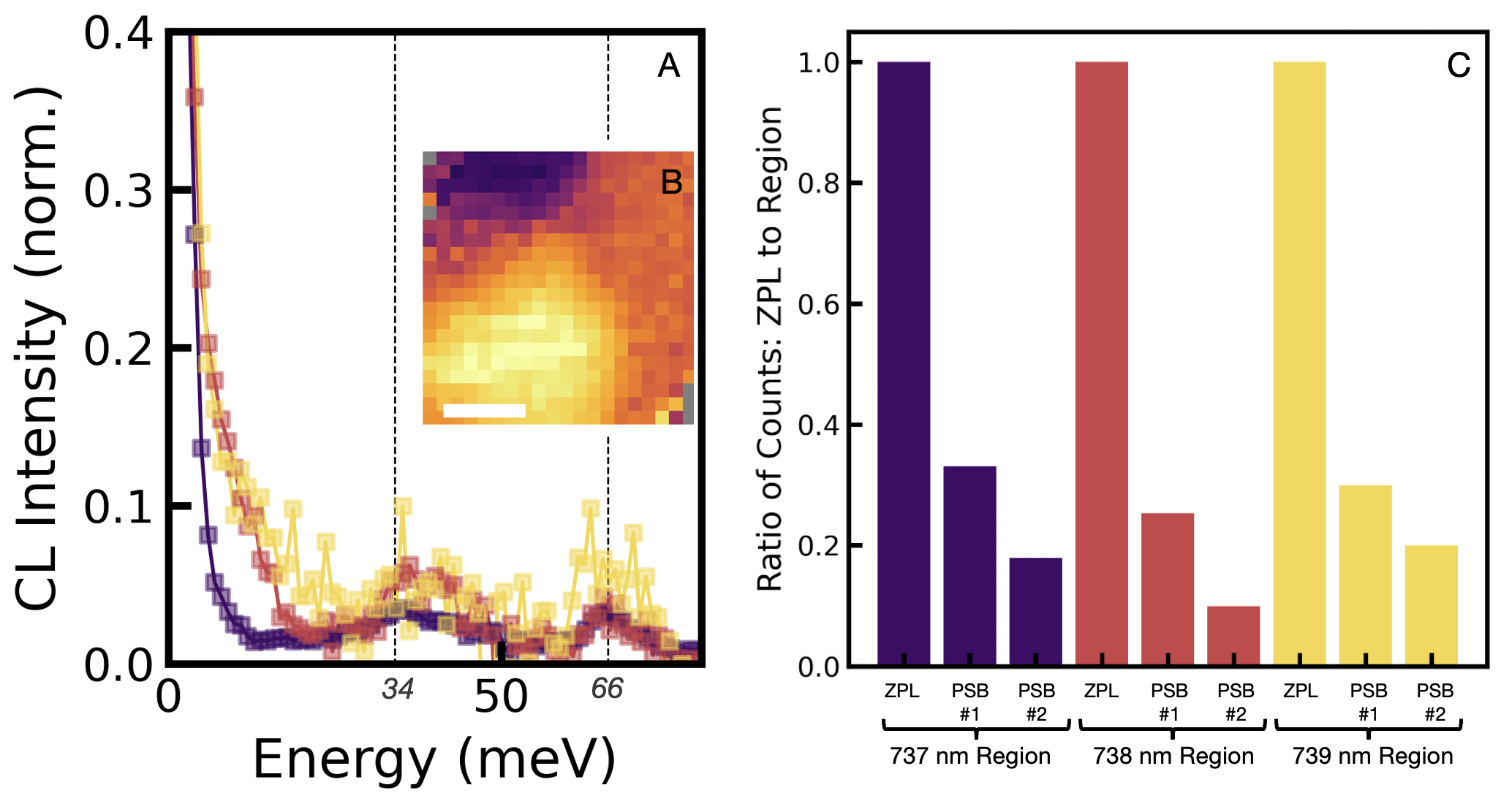}
\caption{ Phonon side band analysis of the particle from Figure 3(e). (a) Normalized phonon side band plots from three different regions (737nm region, 738 nm region, and 739nm region) in map shown in the inset labelled (b). (b) Map of central wavelengths of SiV$^-$ emission. (c) Bar graph of the ratio of integrated intensity found in the PSB, to that found in the ZPL, for the ZPL and the 2 PSB modes: PSB \#1 at 34 meV, and PSB \#2 at 66 meV.}
\label{SI:PSB}
\end{figure}

CL intensity within the ZPL should also depend on electron-phonon coupling. To check if ZPL intensity change is a manifestation of a changing Debye-Waller factor, we check the ratio of counts between the ZPL and two low energy phonon modes at  34 and 66 meV, and find consistent electron-phonon coupling across the diamond boundaries (figure \ref{SI:PSB}). Although this data is noisy, we are only trying to prove that CL intensity is not shifted into the PSB from the ZPL for the red (738 nm region) and yellow (739 nm region). Since the data does not show a significant increase in the PSB to ZPL ratios for these regions, we can conclude that a change in electron-phonon coupling isn't a major contributor to 738$\pm$3 nm intensity variations across crystal boundaries.

\clearpage

\begin{figure}[!hb]
\centering
\includegraphics[width=90mm]{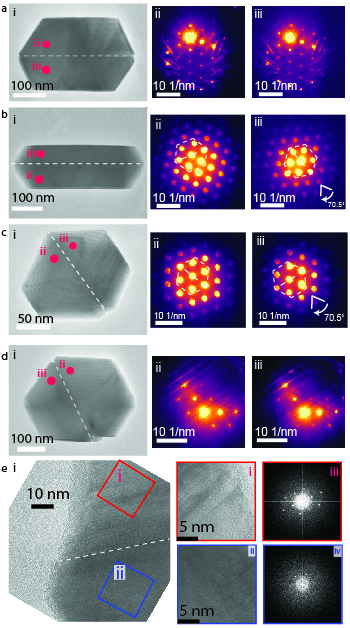}
\caption{(a) i: TEM image. ii-iii: Convergent beam electron diffraction patterns from red circles in i. (b) i: TEM image. ii-iii: Convergent beam electron diffraction patterns from red circles in i. (c) i: TEM image. ii-iii: Convergent beam electron diffraction patterns from red circles in i. (d) i: TEM image. ii-iii: Convergent beam electron diffraction patterns from red circles in i. (e) i: TEM image, i-ii: cropped TEMs from regions labelled i and ii ei. iii-iv: FFTs of cropped TEMs of i and ii.}
\label{SI:Orientation}
\end{figure}

Varying crystal orientation relative to the electron beam can lead to non-stoichiometric EELS loss spectra when characterizing sample composition, the so-called ALCHEMI effect~\cite{brown1999transmission}. Although this effect is significant, CL generation should be less susceptible to electron-beam channeling effects since the bulk plasmon resonance wavelength is on the order of 10nm, much larger than diamond's unit cell.  Nevertheless, to ensure the CL dependency is not an artifact of crystal orientation, CBED patterns can be taken on either side of the nanodiamond boundary, showing that the crystal orientation remains unaltered. This is shown in Figure \ref{SI:Orientation}, where we show that the diamond lattice is oriented consistently across the boundary in question. In the case, of Figure \ref{SI:Orientation}(e) (this is particle Figure 3(a)), we show that the lattice is rotated on either side of the boundary, but still on the $\{110\}$ zone; since the electron beam is symmetric in rotation, these geometries are equivalent, even though the lattice is rotated by 70.5\degree.

Changes in the local density of optical states can alter the radiative efficiency of emission centers, namely the Purcell enhancement~\cite{}. However, if we take particle 10 in Figure 3(a) for example, we see a large decrease in 738$\pm$3 nm intensity across the low energy sigma 3 boundary, within 5.4 nm. Two emitters separated by 5.4 nm (deep subwavelength) will have a virtually identical LDOS, since the sigma 3 boundary won’t alter the refractive index~\cite{}. Therefore we can conclude that the change in 738$\pm$3 nm intensity across this boundary is not due to Purcell enhancement, and the same argument holds for the rest of the particles studied.

Emitted intensity could also fluctuate from the SiV$^{-}$ defect if the electron beam is inducing charge state variations of the defect itself; charge state fluctuations of defects in diamond have been associated with emitter blinking, as well as an emitter being put into a dark state~\cite{bradac_observation_2010,lindner_strongly_2018}. Because the SiV$^{-}$ is negatively charged, the most likely ionization mechanism would be changing the defect from SiV$^{-}$ to SiV$^{0}$, and subsequently shifting the ZPL from 738 nm to 946 nm~\cite{green_neutral_2017}. If such a process is occurring, we would expect to see the appearance of a ZPL at 946nm, which we do not observe.

\clearpage

\begin{figure}[!hb]
\centering
\includegraphics[width=90mm]{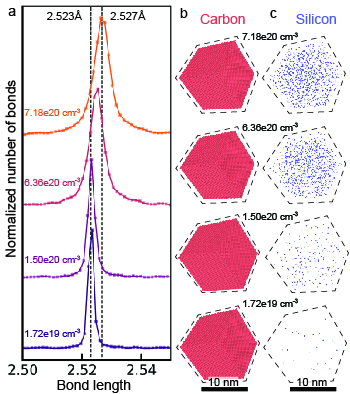}
\caption{LAMMPS simulation of SiV$^-$ dopants in single crystalline nanodiamonds of carbon. (a) Histograms of the 2nd nearest neighbor bond lengths of the diamond tetrahedral, for varying concentrations of the $D_{3d}$ SiV$^-$ defect configuration. (b) the carbon atom positions for each simulation (c) the silicon atom positions for each simulation}
\label{SI:LAMMPS}
\end{figure}

It is possible that discrete changes in SiV defect density across the nanodiamond grainboundaries causes both the intensity change, as well as the shift in ZPL due to internal defect-induced crystal strain. Previously, it has been shown that defect (notably nitrogen and boron defects) implantation into a diamond lattice can be facet-dependent~\cite{wilson_impact_2006,szunerits_raman_nodate,burns_growth-sector_1990}. Intuitively, a large change in the defect density within a crystal lattice will lead to large changes in the lattice's stress state~\cite{anthony_stresses_1995, rossi_micro-raman_1998}. In a diamond lattice, changes in lattice stress states have been experimentally measured across grain boundaries via Raman spectroscopy of the optical phonon~\cite{rossi_micro-raman_1998,vlasov_analysis_1997}, where the 2D defect has been shown to be a boundary of discretely changing vibrational spectra. 
 
Silicon defects in diamond expand the lattice due to Si’s larger diameter. This is confirmed using molecular dynamics simulations in LAMMPS with a Tersoff potential (figure \ref{SI:LAMMPS})~\cite {erhart_analytical_2005}.

\clearpage

\begin{figure}[!hb]
\centering
\includegraphics[width=150mm]{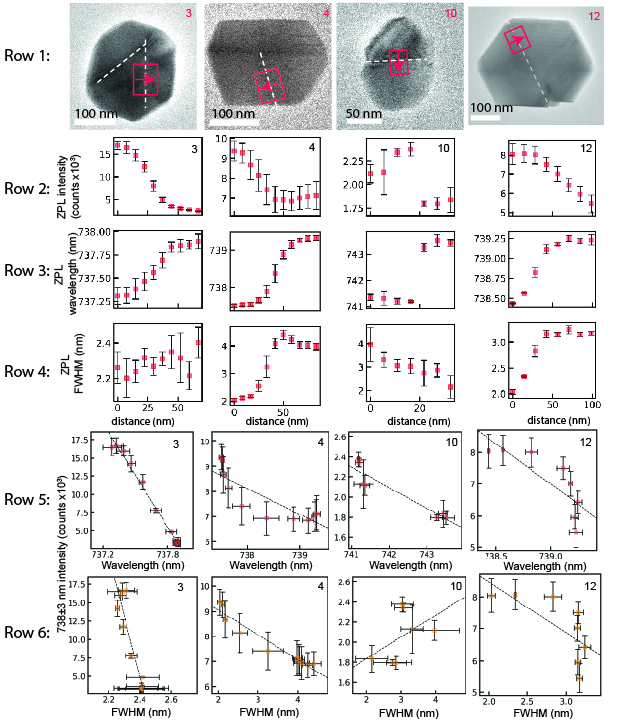}
\caption{Row1: TEM images, Row 2-4: line profiles along red arrows of the ZPL intensity, ZPL wavelength, and the ZPL FWHM, respectively. Row 5: ZPL intensity vs Wavelength, Row 6: ZPL intensity vs FWHM.}
\label{SI:Lineplots}
\end{figure}

\clearpage

\begin{figure}[!hb]
\centering
\includegraphics[width=150mm]{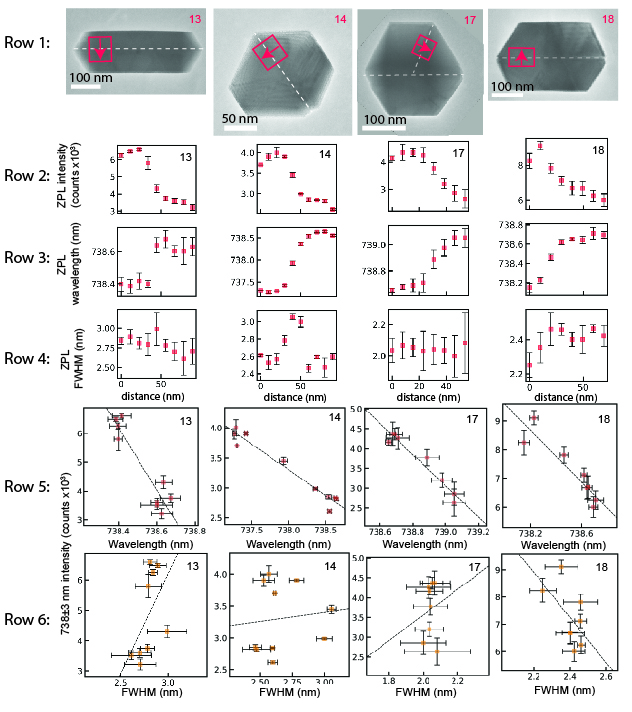}
\caption{Row1: TEM images, Row 2-4: line profiles along red arrows of the ZPL intensity, ZPL wavelength, and the ZPL FWHM, respectively. Row 5: ZPL intensity vs Wavelength, Row 6: ZPL intensity vs FWHM.}
\label{SI:Lineplots2}
\end{figure}

\clearpage

In figures \ref{SI:Lineplots} and \ref{SI:Lineplots2} we show the trends observed at all grain boundaries where we observe heterogeneity. In the first row, we show HRTEM images, where the grain boundaries are delineated with a white dashed line. The red box indicates where data was taken spatially. The data is then binned parallel to the grain boundary (perpendicular to the red arrow). These bins are then averaged and scatter-plotted, where error bars show the standard deviation of the bins. In all cases but one, we see a large decrease (as much as ~70\%) in the CL intensity, as we cross the grain boundary, and this decrease is also accompanied by a redshift in the ZPL (shown in Row 5). The exact trend is not strictly linear (such as particles 4 and 12), although analyzing such trends quantitatively will suffer from emitter convolution. One can imagine that if the grainboundary is perfectly aligned in relation to the electron beam optical axis, the trend should not be linear, but manifest itself as a step function as we cross the domain boundary (such as particle 10, or particle 13). However, if that boundary is angled with respect to the electron beam, as the beam crosses the boundary, both domains can be excited simultaneously, which has the potential to produce linear trends, or the trends shown in particles 4 and 12. Specifically for particle 12, we can see from Figure \ref{SI:p12}(f), that the actual CL spectra appear to show that two emitter ZPLs are being excited at one electron-beam position probe, which is why the FWHM of the fit and the ZPL of the fit appear to vary co-linearly across the boundary; the boundary must be angled significantly with respect to the electron-beam vertical optical axis.

\clearpage

\subsection{Hyperspectral data of small nanodiamonds}

\begin{figure}[!hb]
\centering
\includegraphics[width=130mm]{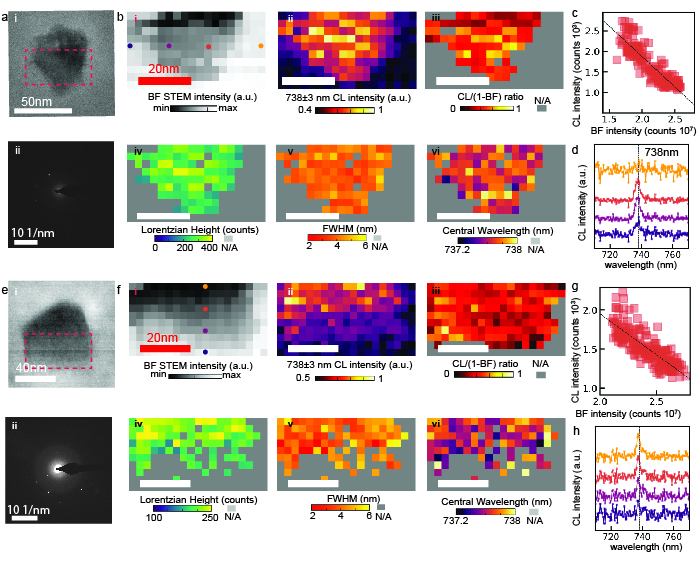}
\caption{2 nanodiamonds below 50 nm in size. (a) i: TEM image ii: Diffraction pattern (b) i-vi STEM intensity, 738$\pm$3 nm intensity, ratio of map ii to map i, where gray pixels are below the STEM threshold, Lorenztian height, FWHM of lorentz fit, Central wavelength of lorentz fit, where gray pixels are below the fit error threshold. (c) 738$\pm$3 nm CL intensity vs BF intensity (scattering map (b)i vs map (b)ii). (d) Point spectra corresponding in color to circles in (b)i. (e)  i: TEM image ii: Diffraction pattern. (f) i-vi STEM intensity, 738$\pm$3 nm intensity, ratio of map ii to map i, where gray pixels are below the STEM threshold, Lorenztian height, FWHM of lorentz fit, Central wavelength of lorentz fit, where gray pixels are below fit error threshold. (g) 738$\pm$3 nm CL intensity vs BF intensity (scattering map (f)i vs map (f)ii). (h) Point spectra corresponding in color to circles in (f)i.}
\label{SI:smalldiamonds}
\end{figure}

We study nanodiamonds as small as 40 nm in diameter as shown in Figure \ref{SI:smalldiamonds}. The hyperspectral data in this figure have pixel sizes of 2.9 nm. These datasets represent the highest resolution in hyperspectral CL mapping we achieved during this study. From the data, it is clear that at this size range, we do not observe any of the heterogeneities that we have now associated with differences in subcrystallite emission. Here, changes in CL emission occur linearly with changes in  approximate particle thickness, which is clear from the scatter plots of CL intensity vs BF STEM intensity, as shown in \ref{SI:smalldiamonds} (c) and (g).

\clearpage

\subsection{Hyperspectral data of heterogeneous, multicrystalline nanodiamonds}

In this section, we show the hyperspectral data that was collected and used for this study from multiple nanodiamonds that weren't shown in the main text. 

\begin{figure}[!hb]
\centering
\includegraphics[width=130mm]{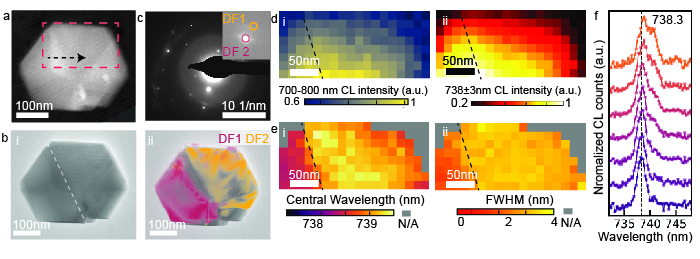}
\caption{(a) DF STEM image (b) i: TEM image, ii: Dark field images overlaid onto TEM image (c) Diffraction pattern. (d) i: 700-800 nm CL intensity ii: 738$\pm$3 nm CL intensity. (e) i: Central wavelength from Lorentz fit ii: FWHM of central wavelength. (f) CL point spectra taken equidistantly along the black dashed line in (a).}
\label{SI:p12}
\end{figure}

\begin{figure}[!hb]
\centering
\includegraphics[width=130mm]{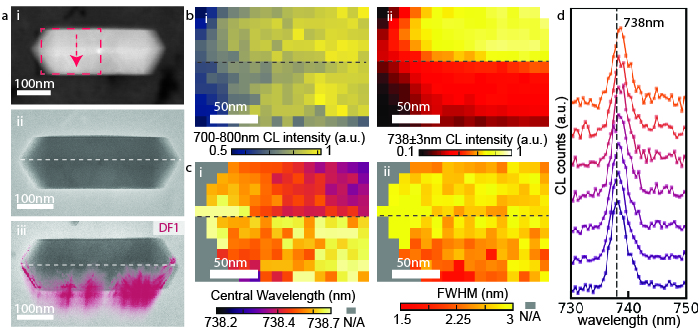}
\caption{(a) i: DF STEM image ii: TEM image, iii: Dark field images overlaid onto TEM image (b) i: 700-800 nm CL intensity ii: 738$\pm$3 nm CL intensity. (c) i: Central wavelength from Lorentz fit ii: FWHM of central wavelength. (d) CL point spectra taken equidistantly along the black dashed line in (a).}
\label{SI:p13}
\end{figure}

\begin{figure}[!hb]
\centering
\includegraphics[width=130mm]{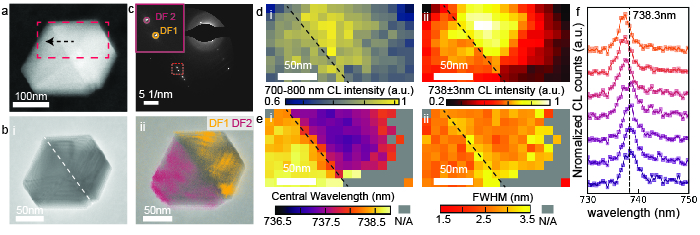}
\caption{(a) DF STEM image (b) i: TEM image, ii: Dark field images overlaid onto TEM image (c) Diffraction pattern. (d) i: 700-800 nm CL intensity ii: 738$\pm$3 nm CL intensity. (e) i: Central wavelength from Lorentz fit ii: FWHM of central wavelength. (f) CL point spectra taken equidistantly along the black dashed line in (a).}
\label{SI:p14}
\end{figure}

\begin{figure}[!hb]
\centering
\includegraphics[width=130mm]{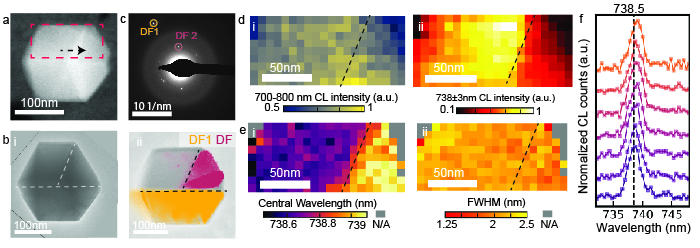}
\caption{(a) DF STEM image (b) i: TEM image, ii: Dark field images overlaid onto TEM image (c) Diffraction pattern. (d) i: 700-800 nm CL intensity ii: 738$\pm$3 nm CL intensity. (e) i: Central wavelength from Lorentz fit ii: FWHM of central wavelength. (f) CL point spectra taken equidistantly along the black dashed line in (a).}
\label{SI:p17}
\end{figure}

\begin{figure}[!hb]
\centering
\includegraphics[width=130mm]{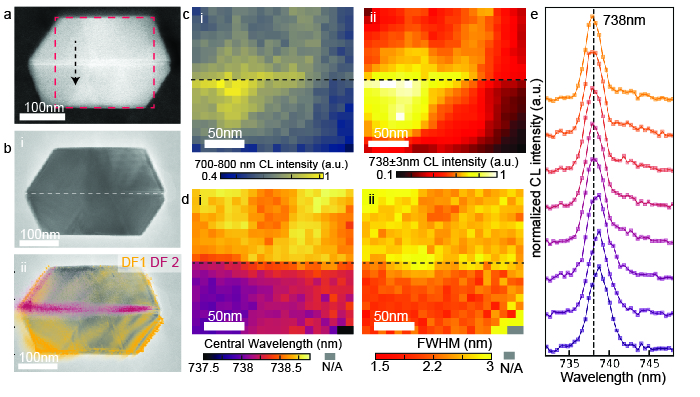}
\caption{(a) DF STEM image (b) i: TEM image, ii: Dark field images overlaid onto TEM image (c) i: 700-800 nm CL intensity ii: 738$\pm$3 nm CL intensity. (d) i: Central wavelength from Lorentz fit ii: FWHM of central wavelength. (e) CL point spectra taken equidistantly along the black dashed line in (a).}
\label{SI:p18}
\end{figure}

\begin{figure}[!hb]
\centering
\includegraphics[width=130mm]{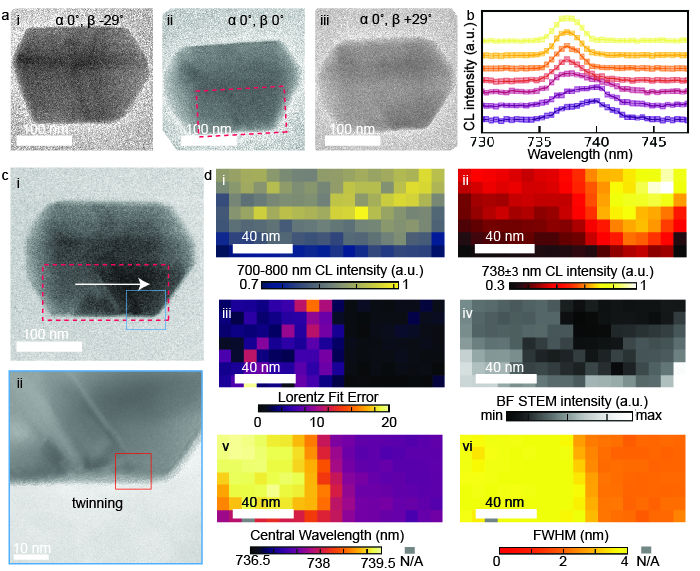}
\caption{(a) i-iii: TEM images at different samples tilts. (b) CL point spectra taken equidistantly along the white line in (c)i. (c) i: BF STEM image ii: TEM image of blue box in i (d) i: 700-800 nm CL intensity ii: 738$\pm$3 nm CL intensity iii: Lorentz error iv: BF STEM intensity v: Central wavelength from Lorentz fit vi: FWHM of central wavelength.}
\label{SI:p4}
\end{figure}

\clearpage

\subsection{Structural analysis of nanodiamonds}

In this section we perform structural analysis on particles shown in Figure 3.

\begin{figure}[!hb]
\centering
\includegraphics[width=110mm]{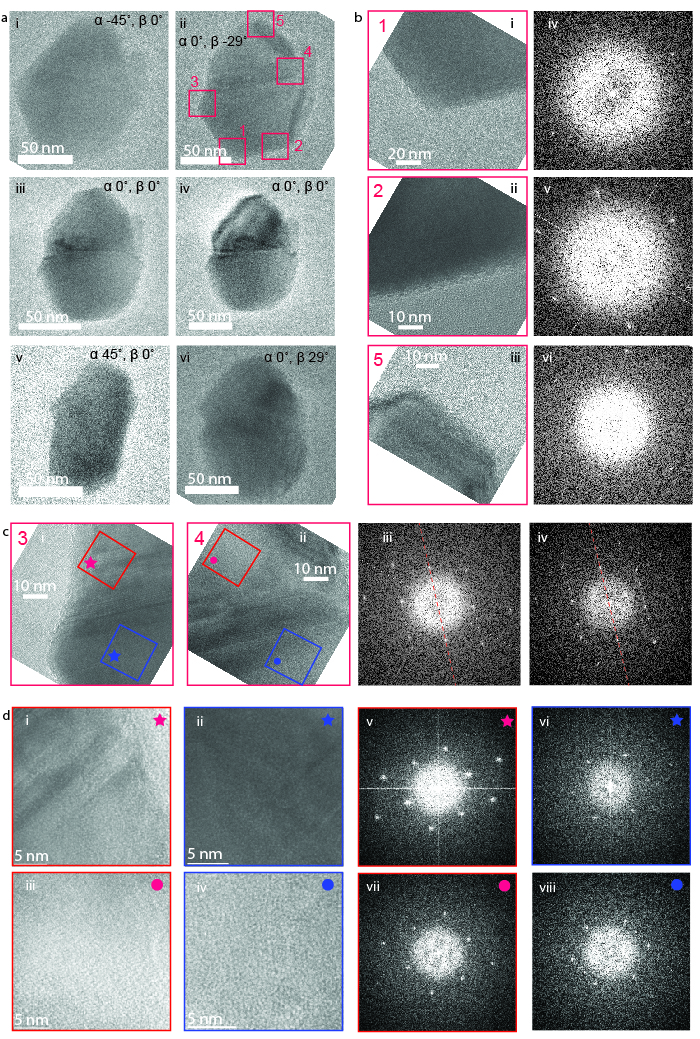}
\caption{(a)i-vi TEM images at different sample tilts (b) i-iii Cropped TEM images of boxes corresponding to (a)ii. iii-vi FFTs of i-iii, respectively. (c) i-ii Cropped TEM images of boxes corresponding to (a)ii, iii-iv FFTs of i-ii, respectively. (d) i-iv cropped TEMs corresponding to ci and cii, indicated by red and blue stars as well as red and blue circles. v-viii FFTs of i-iv, respectively}
\label{SI:p10structure}
\end{figure}

\begin{figure}[!hb]
\centering
\includegraphics[width=130mm]{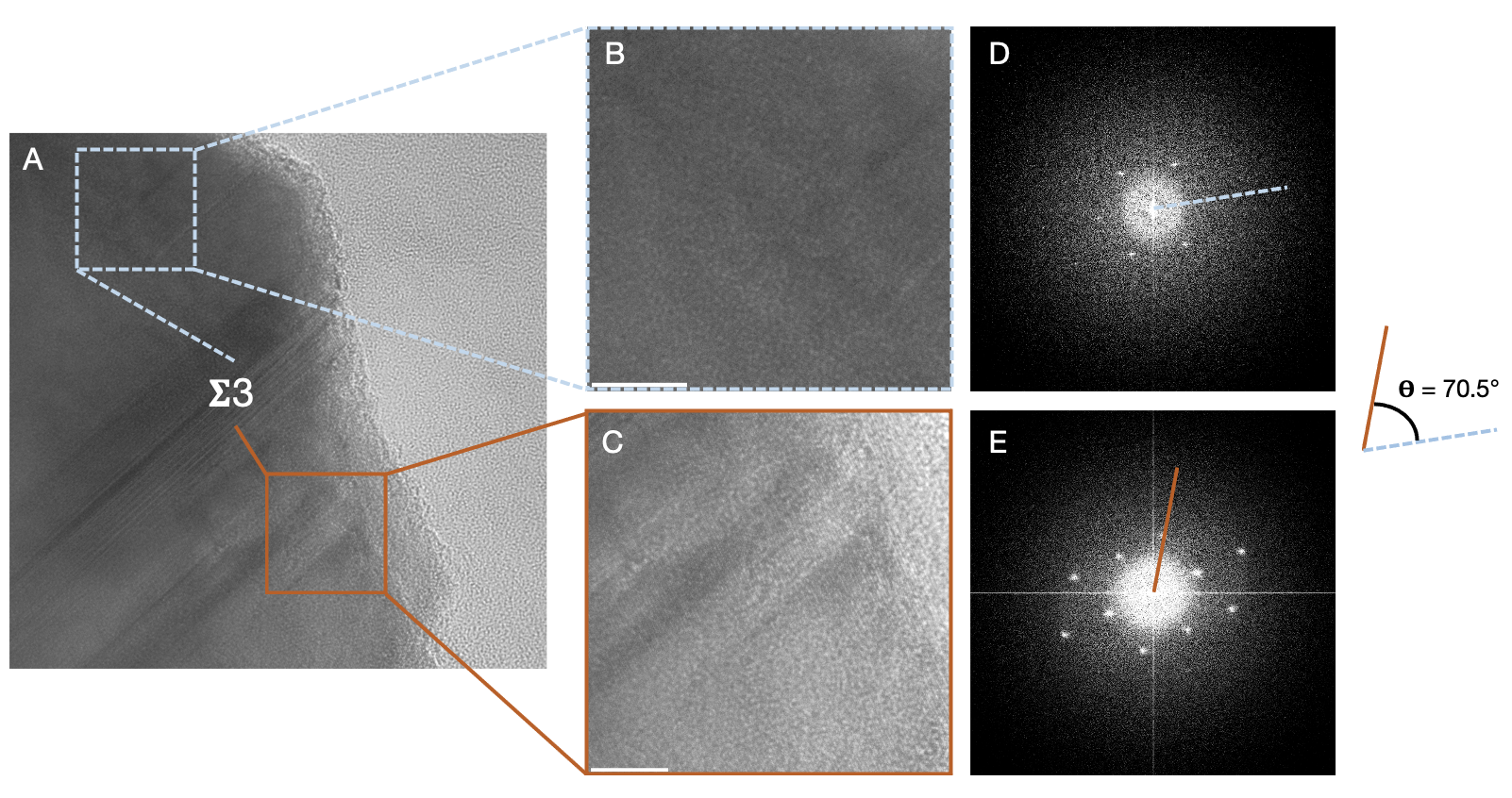}
\caption{ (a) TEM image (b)-(c) cropped TEM images corresponding to squares in (a), (d)-(e) FFTs of (b) and (c), respectively. graphic shows angle difference between the two FFTs of (b) and (c).}
\label{SI:matti_twin}
\end{figure}

By cropping an image of the boundary observed in the particle, we can produce single crystalline FFT patterns on both the top and bottom of the boundary, that are rotated with respect to each other. This is shown in figure \ref{SI:matti_twin}. The angle of this rotation is specific to each type of twin boundary within the diamond cubic lattice. Between the two FFTs, we measure an angle of approximately 70.5\degree. Therefore, the multiple twinned boundary running down the center of this particle is a $\Sigma$3, the lowest energy twin boundary in a diamond lattice~\cite{derjaguin1975structure}. The bonding across at $\Sigma$3 boundary preserves the tetrahedral coordination, thereby maintaining nearest neighbor proximity and making the 2D defect entropic in nature~\cite{liu2009microtexture}. Because of this, one would not expect a multiply twinned $\Sigma$3 boundary to be the source of strain in a diamond lattice~\cite{}. This supports the theory that strain is being produced via another mechanism.

\clearpage

\begin{figure}[!hb]
\centering
\includegraphics[width=130mm]{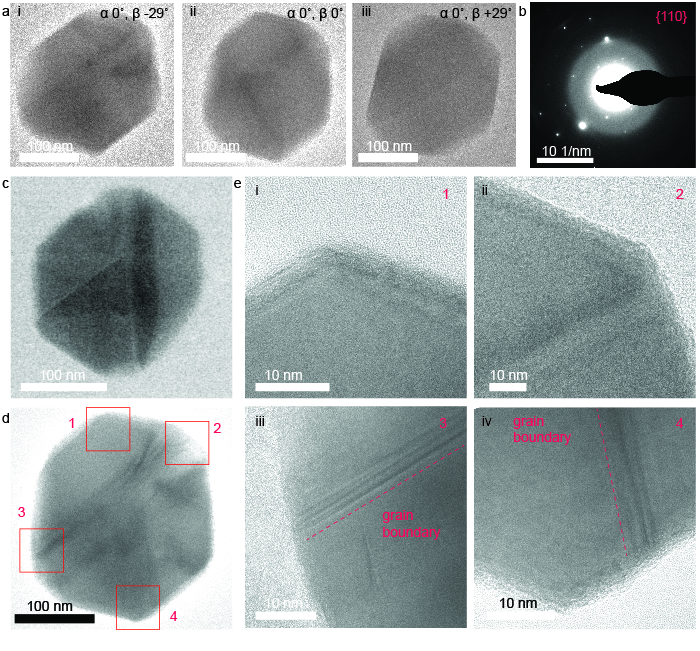}
\caption{ Particle in Figure 3e. (a) i-iii TEM images at different sample tilts (b) Diffraction pattern (c) BF STEM image (d) TEM image (e) i-iv cropped TEM images corresponding to read squares in (d).}
\label{SI:p3_structure}
\end{figure}

In Figure \ref{SI:p3_structure}, we show tilts, the diffraction pattern, as well as HRTEMs of the particle found in Figure 3(e). Because the diffraction pattern in Figure \ref{SI:p3_structure}(b) appears to be single crystalline, it is most likely that the boundaries found in these particles are microtwinned and the twinning results in no rotation between the three main crystallites.

\clearpage

\subsection{Strain analysis on multiple nanodiamonds}

We perform strain analysis on multiple particles with varying geometries. A majority of the particles analyzed show consistent behavior between the local strain state and local optical properties.

\clearpage

\begin{figure}[H]
\begin{center}
\includegraphics[width=115mm]{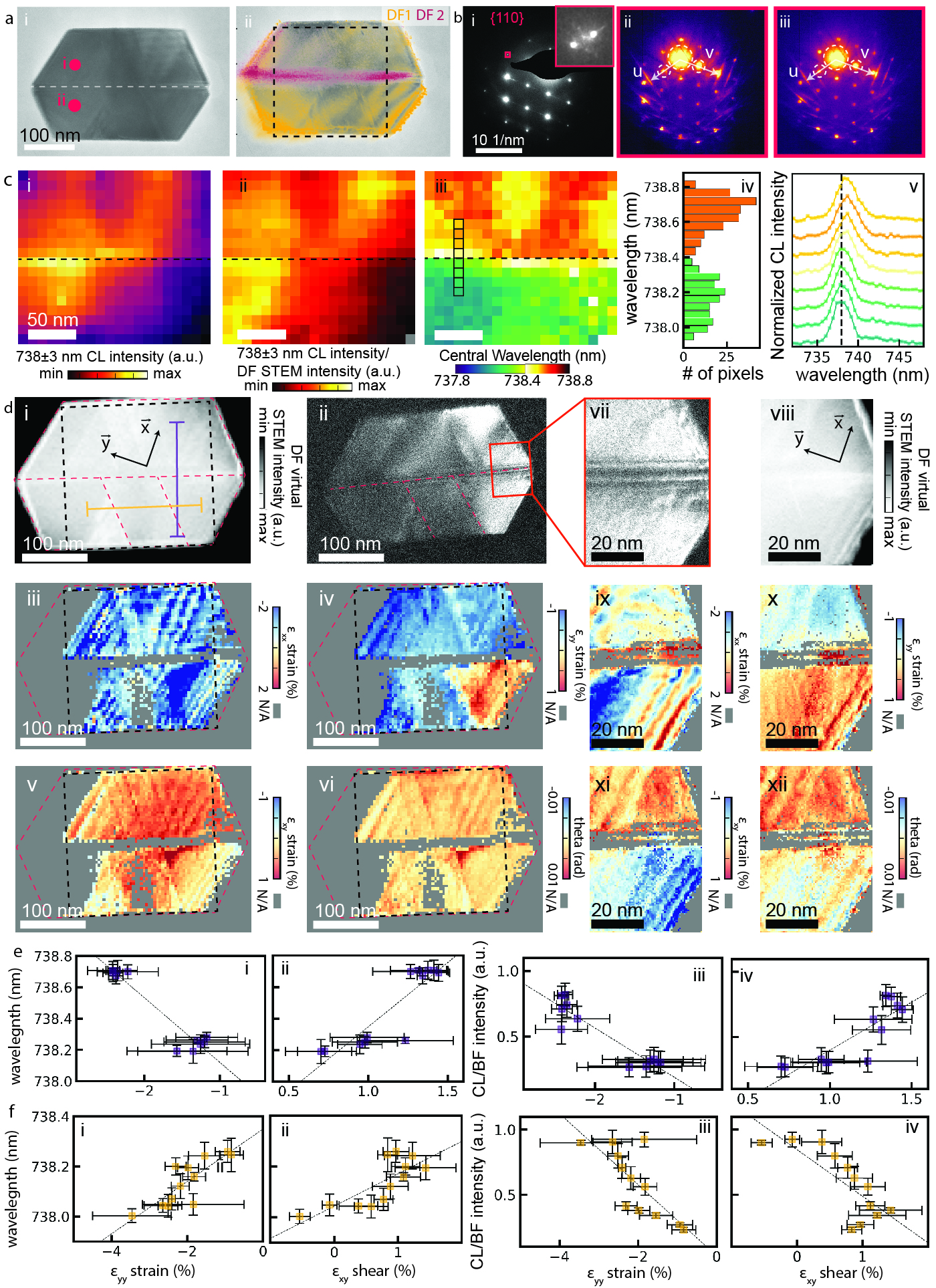}
\end{center}
\caption{ \textbf{SiV$^-$ optical properties correlate with strain  at the nanoscale} (a) i TEM image ii: TEM image with two dark field images overlayed (b) i: whole particle diffraction pattern, inset is a zoom in on one diffraction point boxed in red ii-iii: CBED patterns at red dots in (a). (c) 2D hyperspectral maps taken at black dashed box in aii, i 738$\pm$ 3 nm summed intensity, ii FWHM of Lorentz fit, iii: Central wavelength of Lorentz fit. iv: rows averaged central wavelength, v: histogram of central wavelengths, vi: CL point spectra taken  at black boxes in iii and corresponding in color. (d) i: Virtual DF STEM image, ii: Dark field TEM image, iii-vi: $\epsilon_{xx}$ strain, $\epsilon_{yy}$ strain, $\epsilon_{xy}$ shear, rotation. viii DF virtual image on of red box in ii, viii-xi: $\epsilon_{xx}$ strain, $\epsilon_{yy}$ strain, $\epsilon_{xy}$ shear, rotation. (e) i-ii: ZPL wavelength vs $\epsilon_{yy}$ strain and $\epsilon_{xy}$ shear. iii-iv: Normalized ZPL intensity vs $\epsilon_{yy}$ strain and $\epsilon_{xy}$ shear, for CL data taken along purple line in di that crosses the middle boundary of the particle. (f) i-ii: ZPL wavelength vs $\epsilon_{yy}$ strain and $\epsilon_{xy}$ shear. iii-iv: Normalized ZPL intensity vs $\epsilon_{yy}$ strain and $\epsilon_{xy}$ shear, for CL data taken along orange line in di, that remains in the bottom crystallite.}
\label{SI:p8Strain}
\end{figure}

In  Figure \ref{SI:p8Strain} ci-iii, we show the 738$\pm$3 nm CL counts, 738$\pm$3 nm CL counts normalized by STEM counts, and the ZPL central wavelength, respectively, corresponding in space to the dashed  black box in aii. As seen by the histogram in cv, the nanodiamond is dominated by two groups of emission centered about 738.7 and 738.2 nm, which come from the top and bottom crystallite, respectively. Although the majority of ZPL heterogeneity is due to the differences in top and bottom crystallite, we see that there are still small variations within individual crystallites. For example, in the bottom crystallite, we can see that the emission red shifts towards the right side of the crystallite, and simultaneously decreases in intensity.

Dark field imaging confirms the nanodiamond is separated into two crystallites, by a horizontal boundary, as seen in aii. Additionally, we can identify this grain boundary via the dark field contrast stripes in dii. The whole particle diffraction pattern on the \{110\} zone axis
exhibits fine diffracted streaks running between diffraction points, indicating pronounced twinning of the particle   (Figure \ref{SI:p8Strain} bi).~\cite{walmsley_transmission_1983} CBEDs taken within the top and bottom crystallites shown in bii-iii ( corresponding to red dots in ai), confirm zone-axis consistency across crystallites, which allows for comparative lattice strain analysis.

4D STEM data sets were taken of the entire particle; a virtual Dark Field STEM image is produced from this data set and shown in di, which delineates where strain can be analyzed. We perform strain analysis of the $\epsilon_{xx}$ and $\epsilon_{yy}$ strain, the $\epsilon_{xy}$ shear,  and $theta$ rotation for the entire particle (diii-vi), as well as the particle tip closer to the grain boundary (ix-xii). These maps are produced in the $\overrightarrow{x}$,$\overrightarrow{y}$ basis shown in di, which is rotated by 70.5\degree\ with respect to the image axes. The middle fault separates the crystallite into distinctly strained regions, where the bottom crystallite is expanded and rotated compared to the top (Figure \ref{SI:p8Strain}dx specifically).

\begin{figure}[H]
    \begin{center}
    \includegraphics[width=130mm]{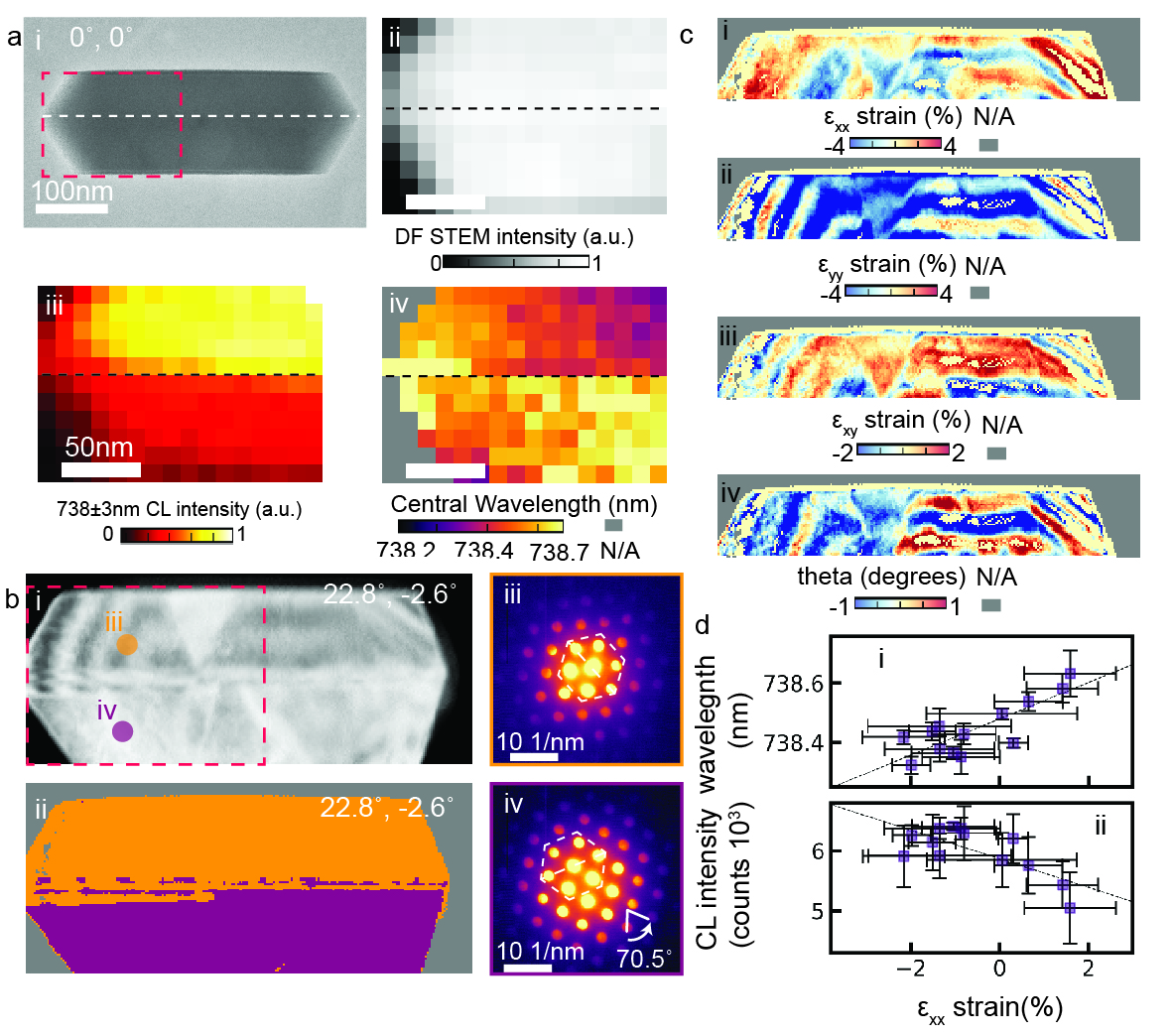}
    \end{center}
    \caption{ \textbf{SiV$^-$ optical properties correlate with strain  at the nanoscale} (a) i TEM image ii: DF stem counts, iii: 738$\pm$3 nm summed intensity,  iv: Central wavelength of Lorentz fit (b) 4D STEM data i: Virtual DF STEM image, ii: Twin crystallite classification image, iii-iv: CBED patterns taken from points in i, corresponding in color. (c) Strain maps from the 4D STEM dataset of the top crystallite i-iv: $\epsilon_{xx}$ strain, $\epsilon_{yy}$ strain, $\epsilon_{xy}$ shear, rotation. (d) i: ZPL wavelength vs $\epsilon_{xx}$ strain and  ii: ZPL intensity vs $\epsilon_{xx}$ strain.}
    \label{SI:p3newStrain}
\end{figure}

\newpage

\begin{figure}[H]
\begin{center}
\includegraphics[width=130mm]{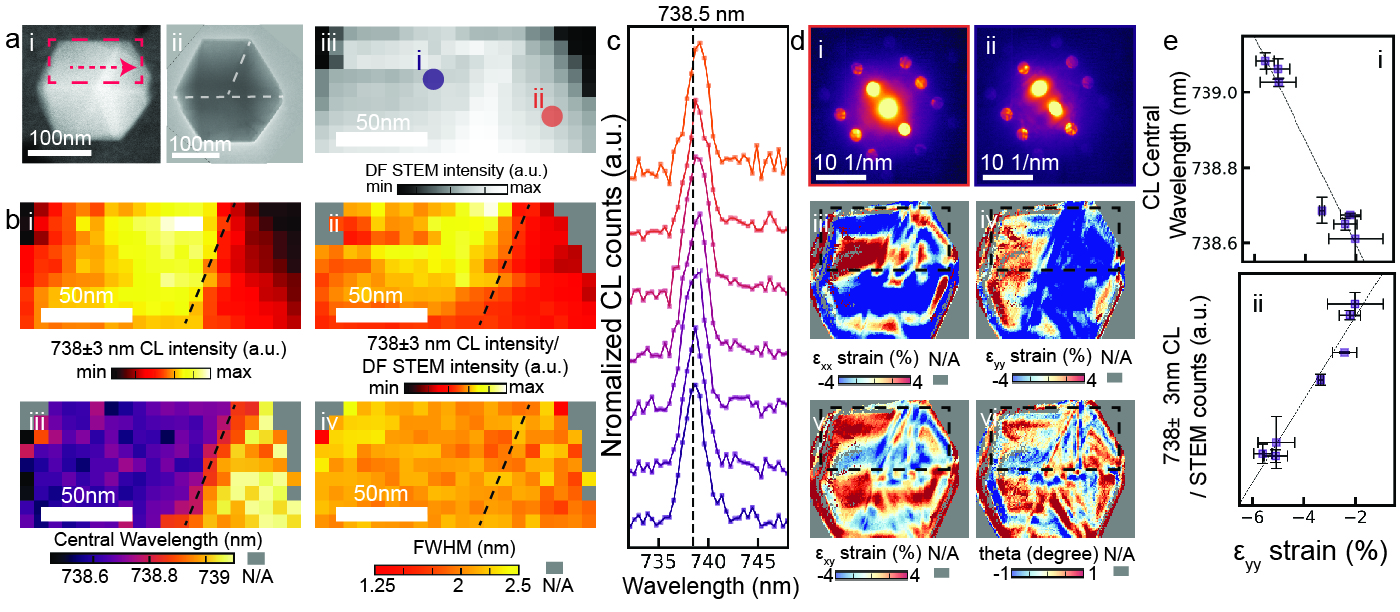}
\end{center}
\caption{ \textbf{SiV$^-$ optical properties correlate with strain  at the nanoscale}. (a) i-ii STEM and TEM images, iii DF STEM counts. (b) i-iv: 738$\pm$3 nm count, 738$\pm$3 nm counts normalized by DF stem counts, Central wavelength of lorentzian fit to 738nm emission, FWHM of lorentzian fit to 738nm emission.correpsonding in space to the red dashed box in ai. scale bars 50 nm. (c) CL point spectra taken along red dashed arrow in ai.(d) i-ii: CBED patterns taken at points in aiii, corresponding in color. iii-vi: strain $\epsilon_{xx}$, strain $\epsilon_{yy}$, shear $\epsilon_{xy}$, rotation $\theta$. (e) i-ii: Central wavelength vs strain $\epsilon_{yy}$, Normalized 738$\pm$3 nm counts vs $\epsilon_{yy}$. }
\label{SI:p17_new_Strain}
\end{figure}
\newpage

\subsection{EELS analysis of diamond bonding at interfaces}

In this section we perform lineplots of the EELS loss spectra as we cross the multicrystalline domain boundaries. Specifically, we show no change in the bulk plasmon intensity, as well as no changes in the $\it{sp3}$ diamond bonding, for multiple particles studied.

\begin{figure}[!hb]
\centering
\includegraphics[width=130mm]{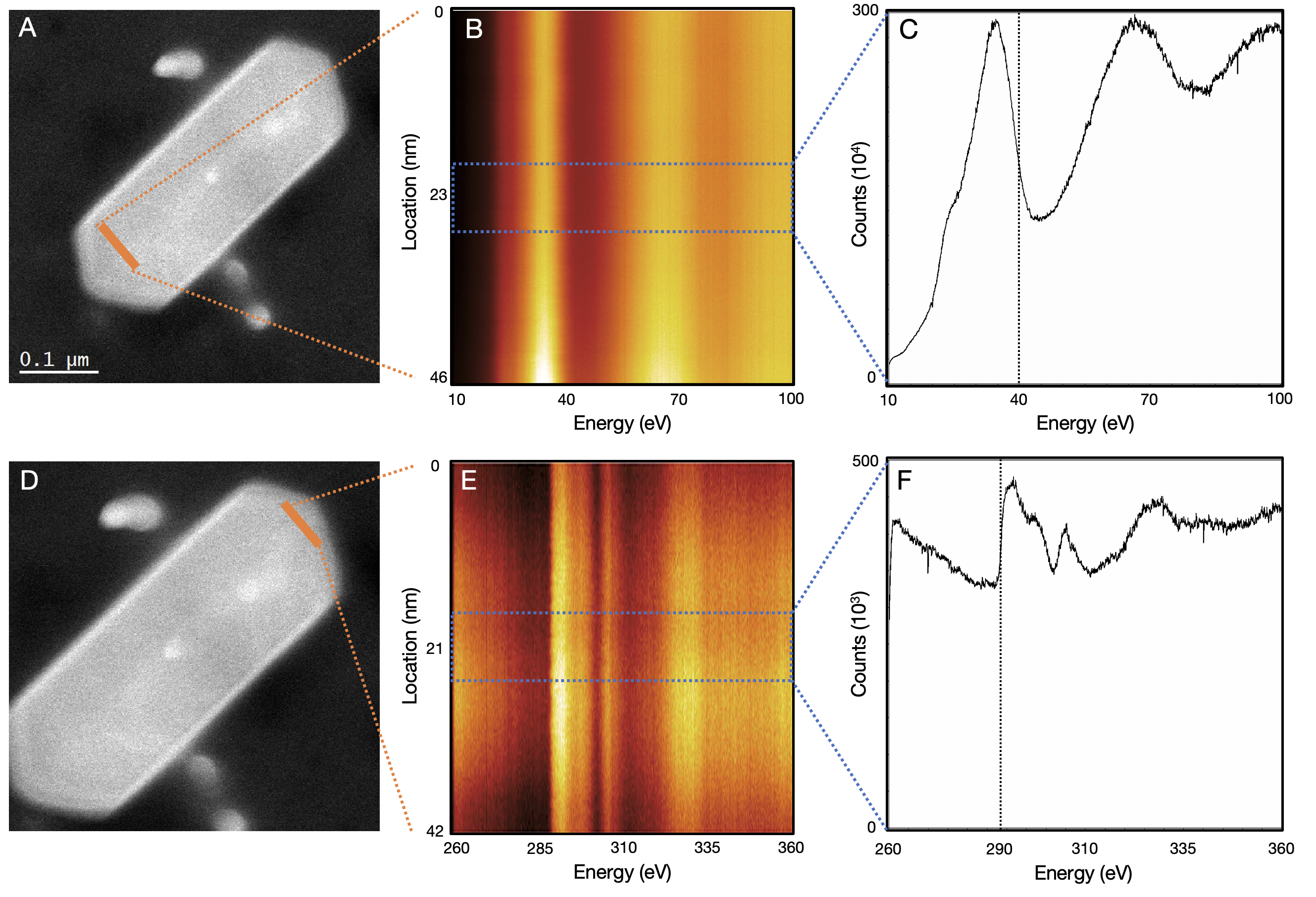}
\caption{Electron Energy Loss Spectroscopy (EELS) line scan analysis of two regions in the particle that is found in Figures \ref{SI:p13} and \ref{SI:p3newStrain}, (a-c) probing the bulk plasmon and (d-e) probing the K-edge. (a) DF STEM image with line scan location, (b) EELS line scan for bulk plasmon energies, (c) composite spectra for the range outlined in blue. (d) DF STEM image with line scan location, (e) EELS line scan for K-edge, (f) composite spectra for the range in blue}
\label{SI:EELS_boundary1}
\end{figure}

\begin{figure}[!hb]
\centering
\includegraphics[width=130mm]{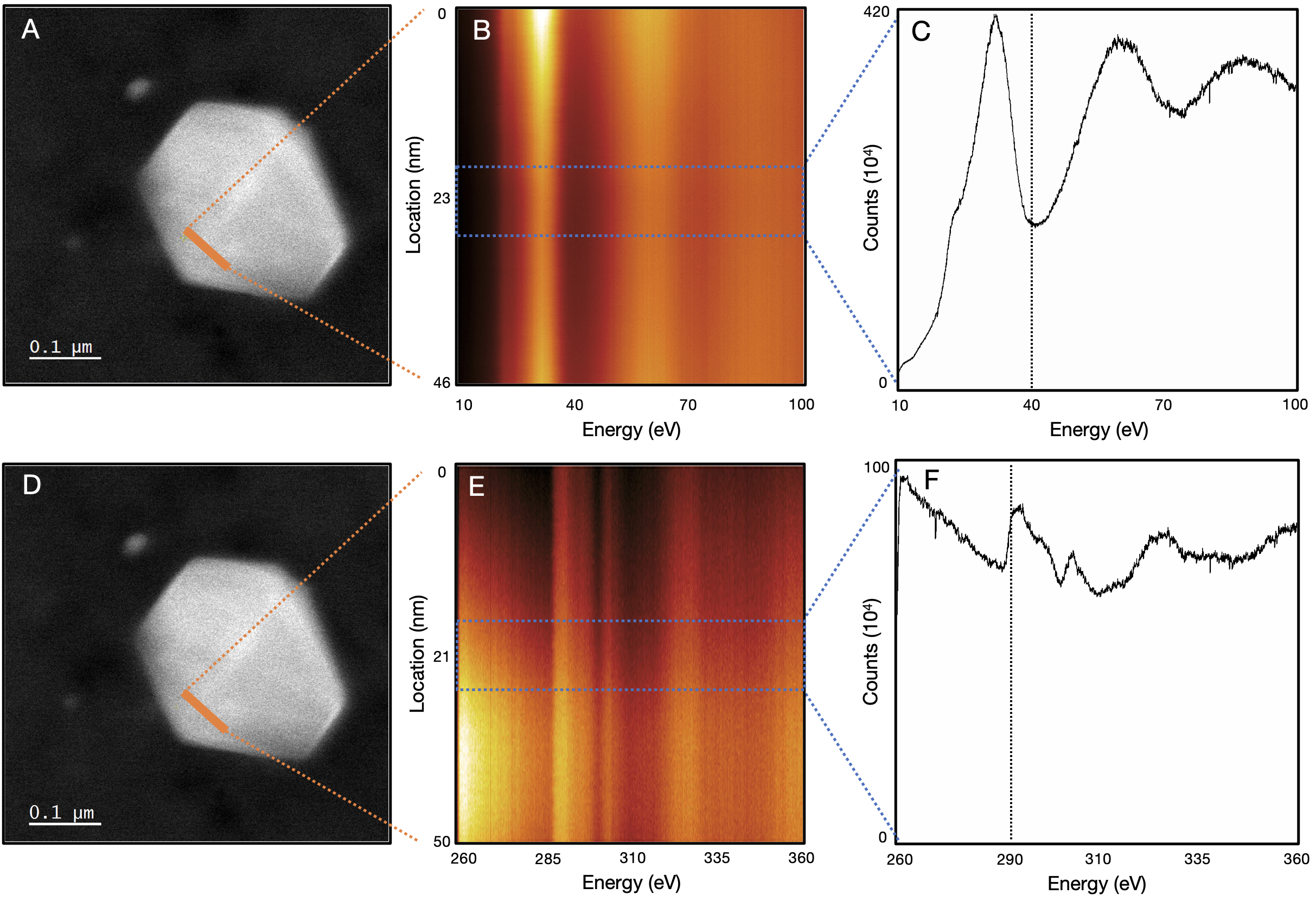}
\caption{Electron Energy Loss Spectroscopy (EELS) line scan analysis of two regions in the particle that is found in Figures 5 and \ref{SI:p14}, (a-c) probing the bulk plasmon and (d-e) probing the K-edge. (a) DF STEM image with line scan location, (b) EELS line scan for bulk plasmon energies, (c) composite spectra for the range outlined in blue. (d) DF STEM image with line scan location, (e) EELS line scan for K-edge, (f) composite spectra for the range in blue}
\label{SI:EELS_boundary2}
\end{figure}

\begin{figure}[!hb]
\centering
\includegraphics[width=130mm]{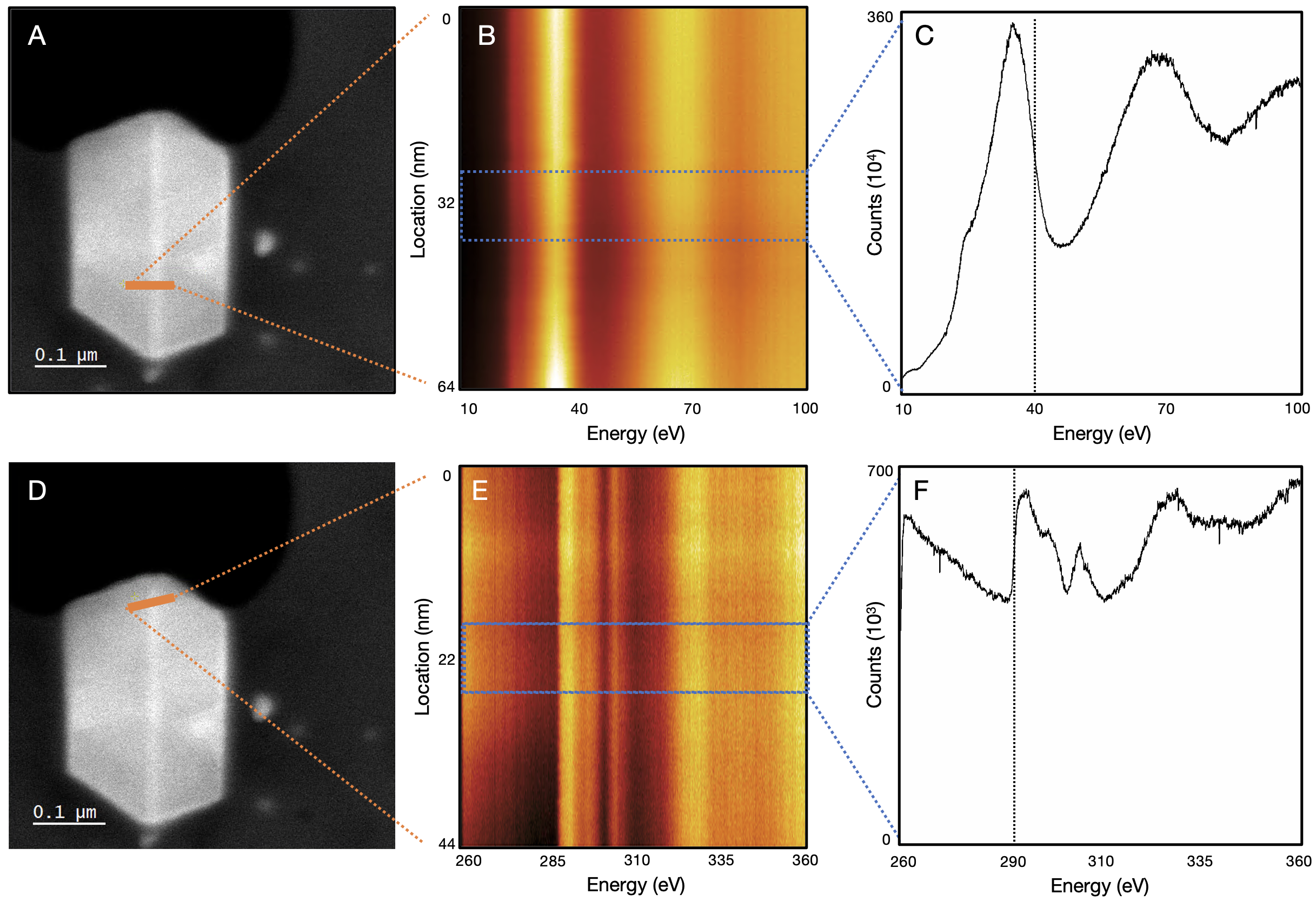}
\caption{Electron Energy Loss Spectroscopy (EELS) line scan analysis of two regions in the particle found in Figures \ref{SI:p18} and \ref{SI:p8Strain}, (a-c) probing the bulk plasmon and (d-e) probing the K-edge. (a) DF STEM image with line scan location, (b) EELS line scan for bulk plasmon energies, (c) composite spectra for the range outlined in blue. (d) DF STEM image with line scan location, (e) EELS line scan for K-edge, (f) composite spectra for the range in blue}
\label{SI:EELS_boundary3}
\end{figure}

\clearpage
\bibliography{Main.bib}

\makeatletter\@input{xx.tex}\makeatother
\makeatletter\@input{output.tex}\makeatother